\documentclass[aps,prc,twocolumn,showpacs,preprintnumbers,superscriptaddress,
                          nofootinbib,float,longbibliography]{revtex4-1}
\usepackage{color}
\usepackage[dvipsnames]{xcolor}
\usepackage{graphicx}
\usepackage{dcolumn}
\usepackage{bm}
\usepackage{appendix}
\usepackage{multirow}
\usepackage[colorlinks=true, pdfstartview=FitV, linkcolor=red,
citecolor=blue, urlcolor=blue]{hyperref}

\usepackage{xcolor}
\usepackage[normalem]{ulem}

\def\be{\begin{equation}}
\def\ee{\end{equation}}
\def\bea{\begin{eqnarray}}
\def\eea{\end{eqnarray}}

\def\kompost{K\o{}MP\o{}ST }
\def\kompostend{K\o{}MP\o{}ST}

\newcommand{\sqrts}{\sqrt{s_\mathrm{NN}}}

\begin{document}


\title{Multi-messenger heavy-ion collision physics}

\author{Charles Gale}
 \affiliation{Department of Physics, McGill University, 3600 University Street, Montreal, QC, Canada H3A 2T8}
\author{Jean-Fran\c{c}ois Paquet}%
\affiliation{Department of Physics, Duke University, Durham, NC 27708, USA}
\author{Bj\"orn Schenke}
\affiliation{Physics Department, Brookhaven National Laboratory, Upton, NY 11973, USA}
\author{Chun Shen}
\affiliation{Department of Physics \& Astronomy, Wayne State University, Detroit, MI 48201, USA}
\affiliation{RIKEN BNL Research Center, Brookhaven National Laboratory, Upton, NY 11973, USA}
%




\date{\today}

\begin{abstract}
This work studies the production of direct photons in relativistic nuclear collisions, along with the production of hadrons. Radiation from the very first instants to the final moments of the evolution is included. The hybrid model used here describes all stages of relativistic heavy-ion collisions. Chronologically, those are an initial state reflecting the collision of nuclei described within the Color Glass Condensate effective theory; a pre-equilibrium phase based on 
non-equilibrium linear response;
relativistic viscous 
hydrodynamics, and a hadronic afterburner. The effect of the pre-equilibrium phase on both photonic and hadronic observables is highlighted for the first time. The potential of photon observables -- spectrum, differential elliptic and triangular flow -- to reveal the chemical equilibration time is studied. Finally, we consider ``small collision systems'', including proton+nucleus collisions and collisions of light nuclei, as probed by hadronic and electromagnetic observables. We demonstrate how photon production can signal the formation of quark-gluon plasma in such small systems.
\end{abstract}

\pacs{Valid PACS appear here}
\maketitle


\section{\label{sec:intro}Introduction}
A great deal is now known about Quantum Chromodynamics (QCD), the theory of the nuclear strong interaction; it is a gauge theory whose degrees of freedom are quarks and gluons \cite{Yang:1954ek}. However, its non-linear nature makes it challenging to solve explicitly, except in some special cases. One of these  involves interactions at large momentum transfers where, because of asymptotic freedom, the smallness of the strong coupling constant admits a perturbative expansion. In that limit, perturbative QCD correctly predicts and interprets a wide class of experimental results \cite{[{See, for example, }][{, and references therein.}]RevModPhys.67.157}. Where perturbative techniques fail, a non-perturbative realization of QCD is accomplished by discretizing the theory on a space-time lattice and by solving numerically the path integral which underpins the theory. Lattice QCD has had remarkable success in its analyses of hadronic observables, both at zero \cite{Bazavov:2009bb} and finite temperature \cite{Ratti:2018ksb}.

Studying the collisions of strongly interacting systems at relativistic energies is currently the only practical means of investigating  finite-temperature and finite-density QCD in terrestrial settings. This is one of the goals of the relativistic heavy-ion program, which is pursued at several large accelerator facilities around the world. In addition to furthering knowledge about QCD under extreme conditions, the heavy-ion program informs our knowledge of strongly interacting systems out of equilibrium. Indeed, one of the breakthroughs in the modeling of relativistic nuclear collisions is the remarkable success of relativistic viscous hydrodynamics, and the realization that the collective behavior observed in heavy-ion final states can reveal departures from equilibrium typically characterized by transport coefficients such as the shear viscosity to entropy density, $\eta/s$, and the bulk viscosity to entropy density ratio, $\zeta/s$ \cite{Gale:2013da}. Theoretical advances such as the ones just described coupled with a vigorous experimental program have been successful in making concrete an early prediction of QCD: the formation of a quark-gluon plasma (QGP) \cite{Baym:2016wox}. The physics of the QGP has now entered a phase of characterization, and most of the contemporary efforts in the field aim to further this goal. 

Several of the observables considered in the analysis of the strongly interacting matter in and out of equilibrium are hadrons, and their final state behavior is often quantified using what has become known as flow analysis \cite{Heinz:2013th}. Quantitatively, in addition to familiar transverse momentum spectra at a given rapidity $y$, the momentum anisotropy is typically quantified in terms of Fourier coefficients as in
\begin{eqnarray}
E\frac{dN}{d^3 p} = \left( \frac{1}{2 \pi p_T} \frac{dN}{d p_T dy}\right)\left[1 + 2 \sum_{n=1}^\infty v_n \cos \left(n(\phi - \Psi_n)\right)\right].\nonumber \\
\end{eqnarray}
The azimuthal angle $\Psi_n$ is a measure of the angle where the measured particles are dominantly produced in the transverse plane. The coefficients $v_n$ are the so-called flow coefficients.

In this work, using the same hybrid approach, we  report on calculations of hadronic and electromagnetic observables, focusing on real photons. The ability to tackle hadronic and electromagnetic observables simultaneously within a single theoretical framework is worthy of emphasis and it is one of the cornerstones of our work: this is multi-messenger heavy-ion physics. Furthermore, in addition to the viscous fluid-dynamical phase that has become the workhorse of relativistic heavy-ion modeling, this work explores the eventually observable consequences of a ``pre-hydrodynamics'' phase. The analyses pursued here use  \kompost \cite{Kurkela:2018wud,Kurkela:2018vqr}, an approach described in more detail in the following section. 

This paper is organized as follows: the next section discusses the hybrid model used to describe the production of hadrons and of photons created during the collision of heavy ions at RHIC and LHC energies.  We then show results obtained for charged hadrons and for the spectra and momentum anisotropies of real photons. In the latter case, the role of chemical equilibrium is examined in some detail. Section \ref{sec:Results} ends with a discussion of ``small systems'', where predictions are made for photons produced in collisions of p+Pb and O+O at the LHC. The paper concludes in Sec.\,\ref{sec:conc}.

\section{\label{sec:Modeling}Modeling the time evolution of relativistic heavy-ion collisions}

In this section, the spacetime modeling of relativistic nuclear collisions is discussed, from the initial stage to the final configurations measured by the detectors. 

\subsection{\label{Kompost}The pre-hydrodynamical phase of relativistic heavy-ion collisions}

The theoretical  modeling of the very early nuclear medium is achieved using the IP-Glasma model. This approach is by now well-established and details are to be found in the original publications \cite{Schenke:2012wb,*Schenke:2012fw}. The IP-Glasma initialization is based on the collision of two color glass condensates, which results in the generation of gluon fields, whose time-evolution is accomplished by solving classical Yang-Mills equations for a proper time duration which we shall label $\tau_{\rm IPG}^{\rm X}$ in this work, where ``X'' is the phase immediately following the IP-Glasma time-evolution. The phase labeled by ``X'' will be one of two, as detailed below. 

One of the goals of this work is to study the effect of considering an out-of-equilibrium era between IP-Glasma and viscous fluid dynamics. The pre-hydrodynamic evolution scheme considered here is that of \kompost \cite{Kurkela:2018wud,Kurkela:2018vqr}. 
In a given time interval $\Delta \tau$, causality sets a limit on the size of the region that can affect conditions at a given spacetime point, $(\tau, {\rm \bf x})$. Taking an average over the causally-connected region, one can split the energy-momentum tensor into a locally homogeneous background in the transverse plane, and perturbations
\begin{eqnarray}
	T^{\mu \nu} \left(\tau_{\rm}, {\bf x}\right)  = \bar{T}^{\mu \nu} \left(\tau_{\rm},{\rm \bf x}\right) + \delta T^{\mu \nu}_{\rm} \left( \tau_{\rm}, {\rm \bf x}\right)\,.
\end{eqnarray}
The evolution of both the background and the perturbation are performed using response functions that scale with $\tau \epsilon^{1/4}/(\eta/s)$, where $\epsilon$ is the energy density, $\tau$ the evolution time and $\eta/s$ the shear viscosity to entropy density ratio. 
In \kompostend, these response functions were calculated using pure-glue QCD kinetic theory, but it is now understood that there is a degree of universality in these functions, both for the transversely-uniform background (see Ref.~\cite{Giacalone:2019ldn} and references therein) and the linear perturbations~\cite{Kamata:2020mka}.

The procedure for obtaining all the elements of the energy-momentum tensor, together with its spacetime dependence is detailed in Refs.~\cite{Kurkela:2018wud,Kurkela:2018vqr}. 
It is worth noting that the scaling properties of the response functions in \kompost reduce considerably its numerical cost, compared to full simulations of the Boltzmann equation. 
However, there are limitations to K\o{}MP\o{}ST, in its current version. First, it is conformal, which inevitably leads to discontinuities when matching to the non-conformal QCD equation of state used in hydrodynamic simulations~\cite{Kurkela:2018vqr,NunesdaSilva:2020bfs}.
Moreover, the lack of bulk viscosity in the conformal case likely generates larger early radial flow.
A second limitation is the linear response approximation which limits its applicability in systems with large fluctuations; this includes small systems such as formed in  proton-nucleus collisions.


\subsection{\label{Hydro}The fluid-dynamical phase of relativistic heavy-ion collisions}


The pre-hydrodynamical ``\kompost phase'' (or directly the IP-Glasma stage) feeds into a fluid-dynamical evolution. This hydrodynamic phase is modelled using \textsc{music} \cite{Schenke:2010nt,Schenke:2010rr,Paquet:2015lta}, a 
relativistic fluid-dynamical evolution which also takes into account transport coefficients such as shear \cite{Schenke:2010rr} and bulk viscosity \cite{Ryu:2015vwa,Paquet:2015lta,Ryu:2017qzn}. The specific shear viscosity parameter\footnote{The shear viscosity per unit of entropy density.} is taken in this work to be $\eta/s = 0.12$, equal to that used in the \kompost phase. To make the connection with earlier calculations of hadron and photon observables using IP-Glasma and its Yang-Mills evolution to initialize \textsc{music}, and to quantitatively assess the changes introduced by the pre-hydrodynamic phase on hadron and photon observables, we shall define and compare the following two scenarios:
\begin{description}
\item[I] IP-Glasma initialization with Yang-Mills propagation for $\tau < \tau_{\rm IPG}^{\rm EKT}$ \footnote{EKT stands here for ``Effective Kinetic Theory''.}, followed by a phase where the energy-momentum tensor is evolved with non-equilibrium linear response
\cite{Kurkela:2018wud,Kurkela:2018vqr}. That \kompost phase is then followed by a viscous fluid-dynamical evolution which begins at $\tau = \tau_{\rm EKT}^{\rm hydro}$.
\item[II] IP-Glasma initialization with Yang-Mills propagation for $\tau_{\rm IPG}^{\rm hydro}$, followed directly by a viscous fluid-dynamical evolution.
\end{description}
In this work we adopt $\tau_{\rm IPG}^{\rm EKT} = 0.1$ fm/$c$, which corresponds to a gluon saturation scale $Q_s \sim 2$ GeV. For the transition from \kompost to hydrodynamics, we use $\tau_{\rm EKT}^{\rm hydro} = 0.8$ fm/$c$ as default, in line with the values used in Ref.\,\cite{Kurkela:2018wud}.
For scenario II, $\tau_{\rm IPG}^{\rm hydro} = 0.4$ fm/$c$ is used, consistent with previous works~\cite{Ryu:2015vwa,Gale:2012rq,Gale:2013da}.
To help comparisons between scenarios I and II, a smaller number of additional calculations are also performed in scenario I with $\tau_{\rm EKT}^{\rm hydro} = 0.4$ fm/$c$.

At these switching times the complete energy-momentum tensor $T^{\mu \nu}$ is passed on to the next stage. Schematically: 
\begin{eqnarray}
T_{\rm IPG}^{\mu \nu} (\tau_{\rm IPG}^{\rm EKT}) &=& T_{\rm EKT}^{\mu \nu} (\tau_{\rm IPG}^{\rm EKT})\nonumber \\
T_{\rm EKT}^{\mu \nu} (\tau_{\rm EKT}^{\rm hydro}) &= &T_{\rm hydro}^{\mu \nu} (\tau_{\rm EKT}^{\rm hydro})
\end{eqnarray}
for scenario I, and 
\begin{eqnarray}
T_{\rm IPG}^{\mu \nu} ({\tau_{\rm IPG}^{\rm hydro}}) &=& T_{\rm hydro}^{\mu \nu} (\tau_{\rm IPG}^{\rm hydro})
\end{eqnarray}
for scenario II. In every scenario, longitudinal boost invariance is assumed. 
The fluid-dynamical phase is followed by a ``hadronic afterburner'', which ensures a dynamic kinetic freezeout, after which the hadrons free-stream to the detectors. The Cooper-Frye prescription \cite{Cooper:1974mv} provides the interface between hydrodynamics and UrQMD \cite{Shen:2014vra}. Viscous corrections follow Ref.~\cite{Schenke:2020mbo}: the 14-moment viscous corrections for the shear $\delta f$ and the Chapman-Enskog form for the bulk viscous corrections \cite{Ryu:2015vwa, Schenke:2020mbo}.
In this work, the switching temperature to the afterburner is set to $T_{\rm switch} = 145$ MeV.

\section{\label{results}Results}
\label{sec:Results}

\subsection{\label{results:hadrons} Hadrons}

\begin{figure}[ht!]
\centering
\includegraphics[width=0.95\linewidth]{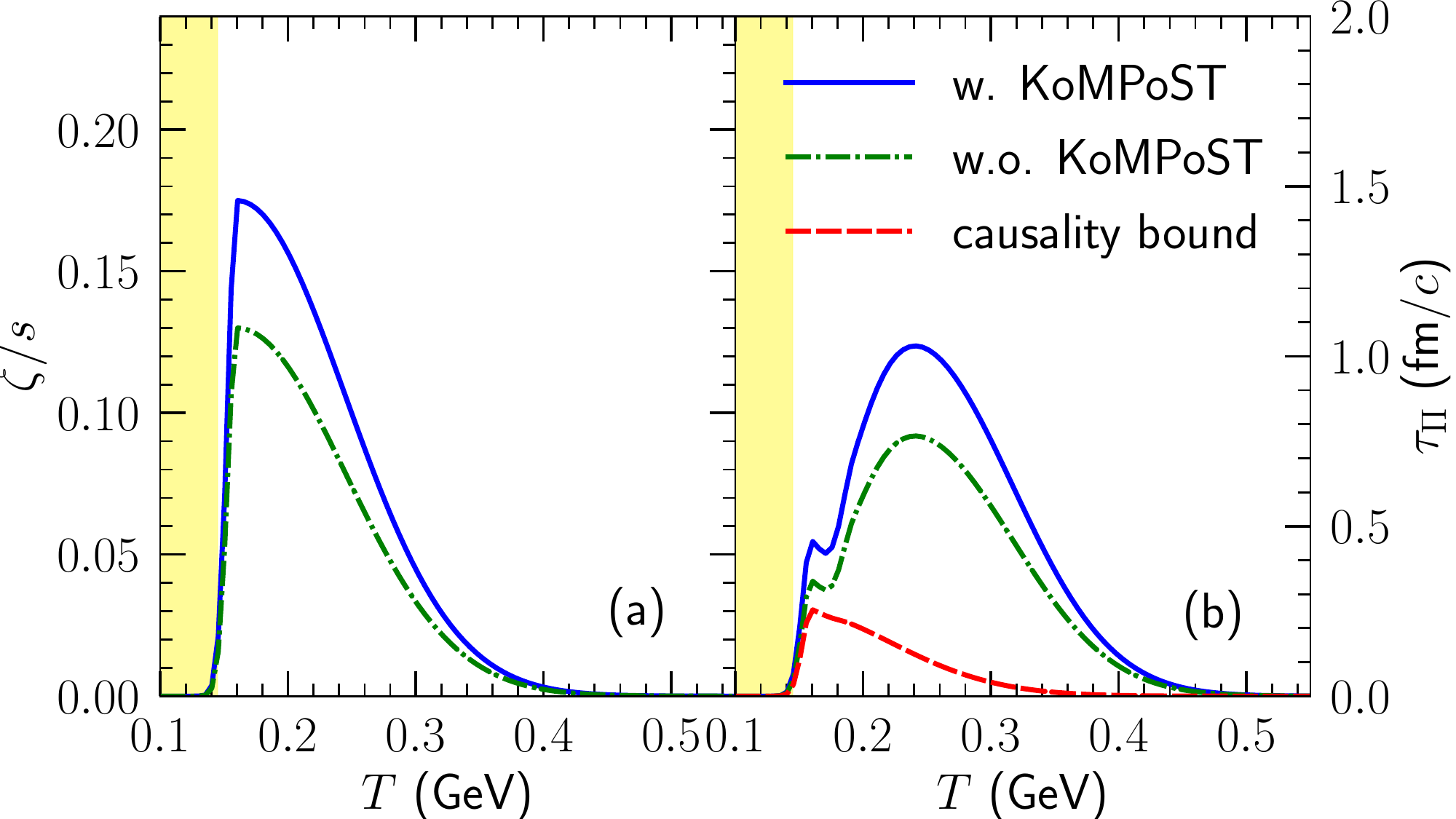}
\caption{(Color online) (a) The bulk viscosity temperature profile consistent with heavy ion data at RHIC and at the LHC, with and without the \kompost phase. Panel (b) shows the corresponding bulk relaxation times and their minimum value required by causality in the linear regime around equilibrium~\cite{Huang:2010sa}.}
\label{fig:Bulk}
\end{figure}

In this subsection, the production of hadronic observables within the hybrid model is reviewed, with focus on explicit studies of the effects of the \kompost phase (by comparing  scenario I and II discussed in the previous section). The results are compared against experimental data gathered at RHIC and at the LHC. 
Two parameters are adjusted to data separately for scenario I and II: the bulk viscosity and the normalization of the energy-momentum tensor.

It is known that the bulk viscosity can 
have a significant effect on hadronic observables~\cite{Ryu:2015vwa}.
The bulk viscosity profile consistent with the hadronic data in scenario I is different from that in scenario II, as shown in Fig.\,\ref{fig:Bulk}. This highlights the first important conclusion of our paper: the presence of a pre-equilibrium phase like \kompost can have a clear influence on the values of the transport coefficients of QCD extracted by interpreting the experimental data. Given the fact that \kompost is currently formulated in a conformal formalism, the quantitative effects may change in a more complete treatment~\cite{NunesdaSilva:2020bfs}. 
Note that in Fig.\,\ref{fig:Bulk}, the shaded region in yellow is where the transport coefficients are handled dynamically by UrQMD. 

The normalization of the initial energy-momentum tensor changes between scenarios I and II because entropy production is different in our kinetic theory and hydrodynamic phases. The \kompost phase in scenario I generates $\sim20\%$ less entropy during the preequilibrium phase compared to the scenario II, which is corrected by a corresponding change in the normalization factor.

\begin{figure}[ht!]
\centering
\includegraphics[width=0.9\linewidth]{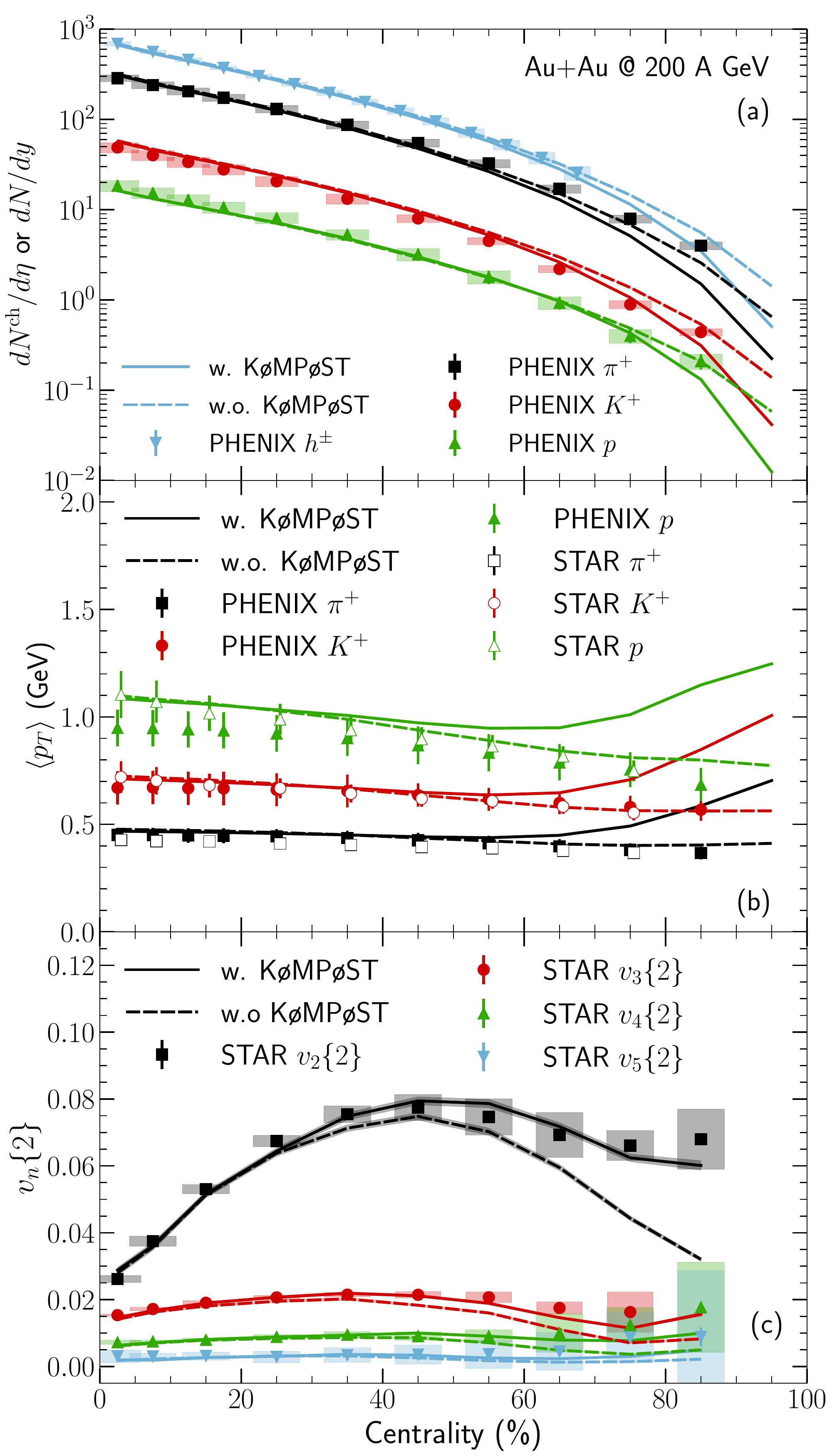}
 \caption{(Color online) For Au+Au collisions at RHIC, the different panels show (a) multiplicity values, (b) average transverse momentum, and (c) momentum anisotropy  coefficients $v_n \{2\}$ as a function of collision centrality, for charged hadrons and identified species. See main text for details. Results obtained with (scenario I) and without (scenario II) \kompost are shown. The data are from Refs.~\cite{Adler:2004zn, Adler:2003cb, Adamczyk:2016exq, Adamczyk:2017hdl}. }
\label{fig:RHIC}
\end{figure}

Starting with Au+Au collisions at the top RHIC energy of 200 A GeV, Fig.\,\ref{fig:RHIC} (a) shows multiplicity as a function of centrality class, for  charged hadrons as well as for individual identified hadrons.  The multiplicities have been integrated over a range of pseudorapidity (or rapidity, for identified hadrons) 
of one unit, centered at the origin.
Once the energy-momentum tensor normalization and the bulk viscosity are tuned to data, the effect of the preequilibrium phase on these is seen to be modest for the shape of centrality dependence
but becomes larger in peripheral collisions starting at $\sim75\%$.
The mean transverse momentum of identified hadrons generated in the relativistic nuclear collisions, $\langle p_T \rangle$, is shown in Fig.\,\ref{fig:RHIC} (b). The transverse momentum in scenario II tracks the experimentally measured values, up to the last centrality bin, whereas that in scenario I deviates from observed values starting at $\sim70\%$ for mesons and somewhat earlier for protons. 
Additional verifications suggest that above is predominantly a consequence of the different hydrodynamics initialization time of \kompost{} and IP-Glasma.
Peripheral collisions generate fluid dynamical phases with a shorter lifetime than those associated with more central collisions. Consequently, the relative time spent in the ``pre-hydrodynamics'' era will appear correspondingly longer.   Since both IP-Glasma and \kompost are conformal \cite{Schenke:2012wb,Kurkela:2018wud}, this leaves a shorter time for the bulk viscosity to generate viscous entropy and reduce the transverse  momentum, as illustrated, for example,  in Refs.~\cite{Ryu:2015vwa, Vujanovic:2019yih, NunesdaSilva:2020bfs}. 

This lack of viscous damping and increase of mean $p_T$ can also be seen in the flow coefficients $v_n\{2\}$ of charged hadrons, obtained from two particle correlations and shown in Fig.\,\ref{fig:RHIC} (c), where the effect of going from scenario I to II grows again with increasing centralities. 
In summary, for the specific choice of hydrodynamic initialization times used in this work, the difference seen between scenario I and II appears to be predominantly from the later hydrodynamic initialization time used with \kompost{} (scenario I). This observation is also made with photons in Section~\ref{results:photons}.

\begin{figure}[t]
	\centering
	\includegraphics[width=0.9\linewidth]{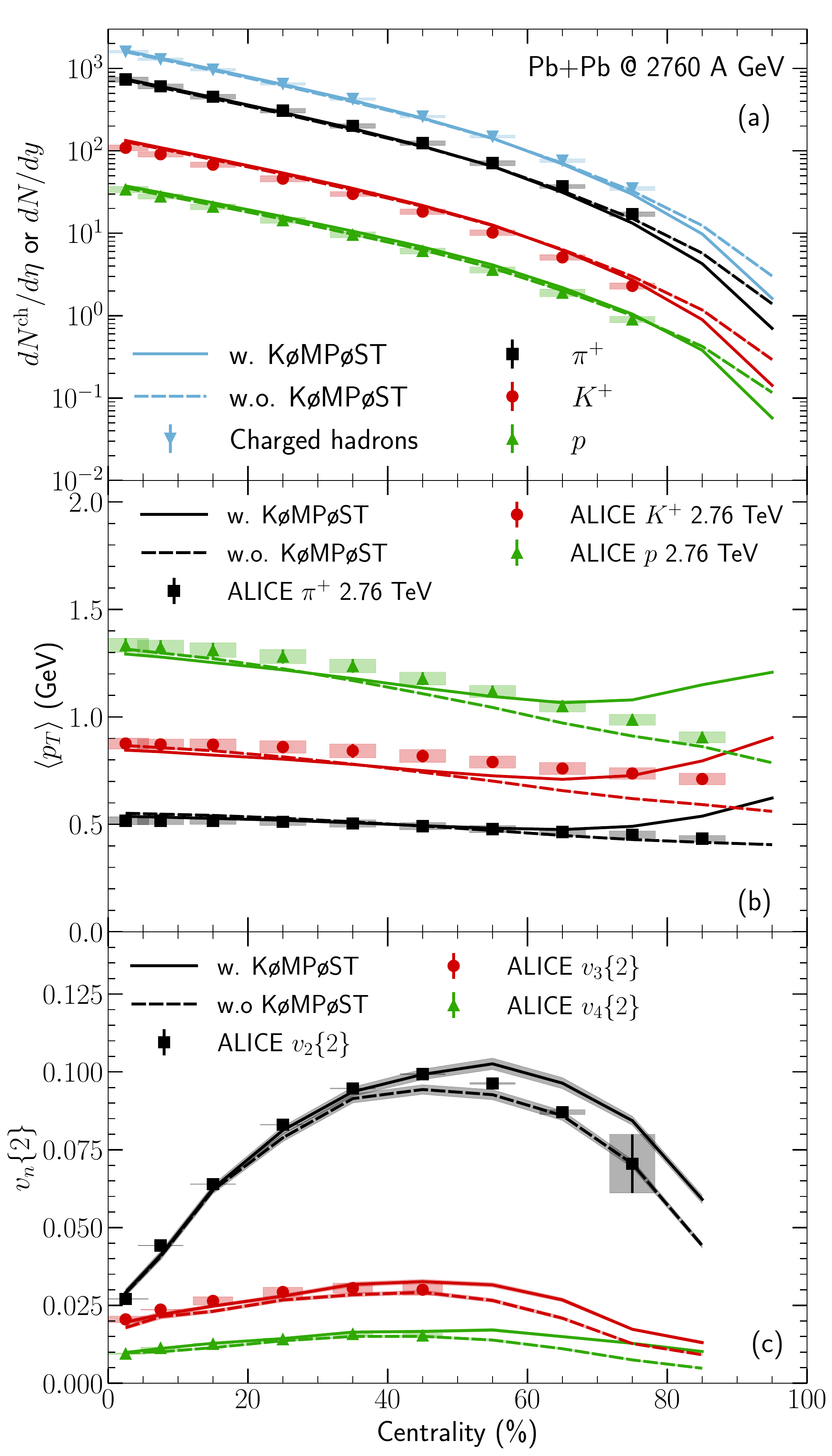}
	\caption{(Color online) For Pb+Pb collisions at the LHC, for an energy of 2760 A GeV. the different panels show (a) multiplicity values, (b) average transverse momentum, and (c) momentum anisotropy  coefficients $v_n \{2\}$ as a function of collision centrality, for charged hadrons and identified species. See main text for details. Results obtained with (scenario I) and without (scenario II) \kompost are shown. The data are from Refs.~\cite{Abelev:2013vea, ALICE:2011ab}. }
	\label{fig:LHC}
\end{figure}

\begin{figure}[ht!]
	\centering
	\includegraphics[width=0.9\linewidth]{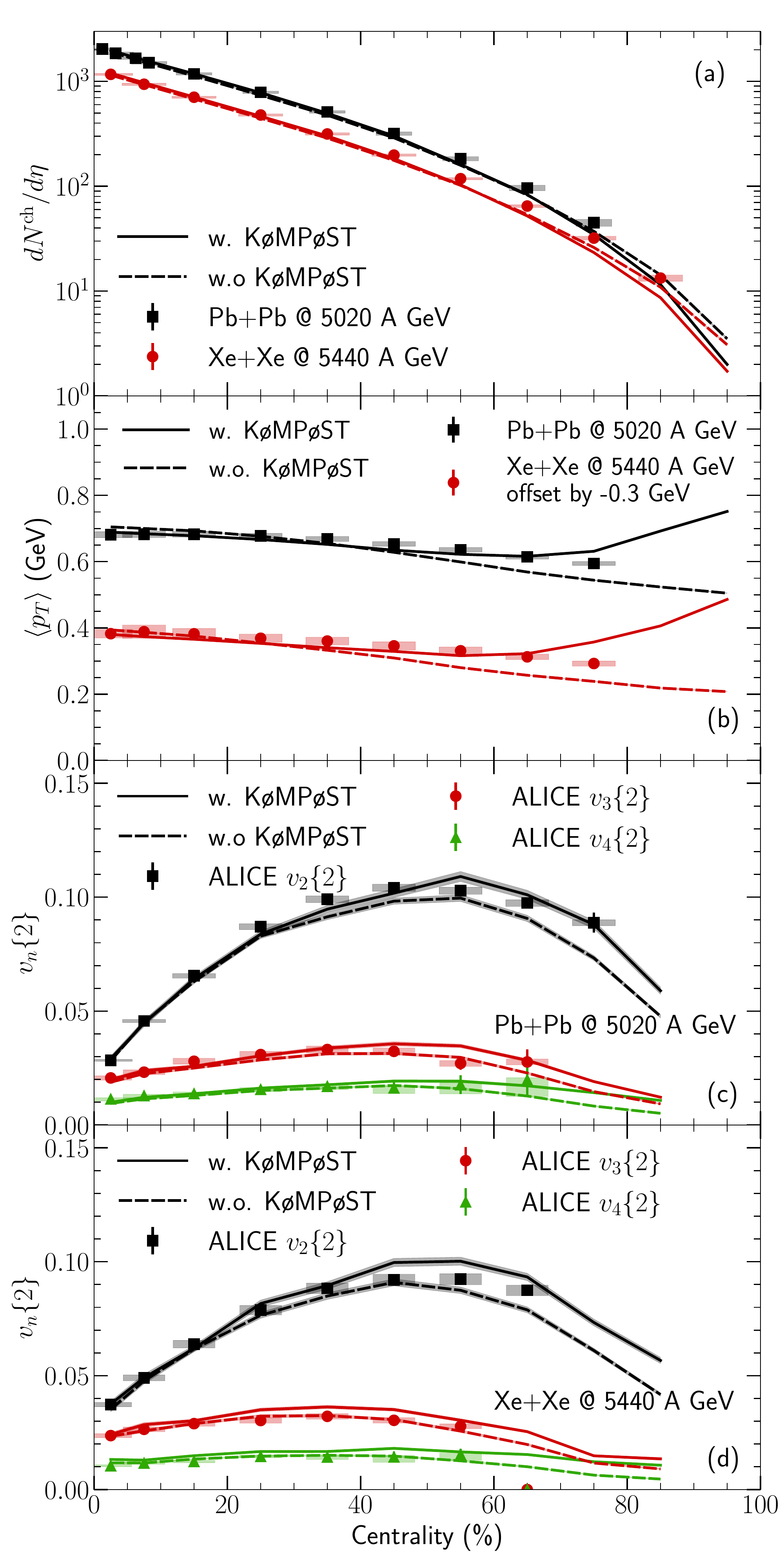}
	\caption{(Color online) Comparison of results obtained for Pb+Pb collisions at an energy of 5020 A GeV, and of $^{129}$Xe+$^{129}$Xe at an energy of 5440 A GeV. The different panels show (a) multiplicity values, (b) average transverse momentum, and (c)+(d) momentum anisotropy  coefficients $v_n \{2\}$ as functions of collision centrality, for charged hadrons and identified particles. See main text for details. Results obtained with (scenario I) and without (scenario II) \kompost are shown. The data are from Refs.~\cite{Adam:2015ptt, Acharya:2018hhy, Acharya:2018eaq, Adam:2016izf, Acharya:2018ihu}. }
	\label{fig:LHC_Xe}
\end{figure}

Concentrating now on conditions prevalent at the Large Hadron Collider, the interpretation of experimental data with our hybrid model is highlighted in Fig.\,\ref{fig:LHC}. 
The same data selection as that for RHIC is shown. Also at the LHC, the effects of the pre-equilibrium phase are seen mostly in peripheral collisions. They are quantitatively comparable to those observed at RHIC energies. Turning to a slightly smaller system at a  higher energy -- $^{129}$Xe + $^{129}$Xe at 5440 A GeV -- yields the results shown in Fig.\,\ref{fig:LHC_Xe}, where we also compare directly to Pb+Pb collisions at 5020 A GeV. We do not observe a dramatic difference in the effects of the pre-equilibrium stage between different systems. Differences appear to set in somewhat earlier for the smaller system, as is qualitatively expected, for the same reason that the effects are larger in peripheral events.

\subsection{\label{results:photons} Photons}

Electromagnetic radiation constitutes another class of observables capable of revealing the nature and behavior of the strongly interacting medium created under the extreme conditions generated by heavy-ion collisions at RHIC and at the LHC. A striking difference between electromagnetic and hadronic observables is that the former will survive unscathed the entire spacetime history of the collision, from the very first interaction to the stage where  final particles are recorded in the detectors. The electromagnetic signal is penetrating, owing to $\alpha/\alpha_{\rm s} \ll 1$. A complete theoretical understanding of the photon (real and virtual) signal will require summing the different components of the photon spectrum, highlighting  the importance of   treating electromagnetic signals with a dynamical approach which is realistic over the entire space-time history of the collision. 
 In this work, the approach outlined in Ref.\,\cite{Paquet:2015lta} is followed, with the exception of an additional stage to be specified shortly. The very first nucleon-nucleon collisions will generate ``prompt photons'', calculated with next-to-leading-order perturbative QCD using INCNLO \cite{Aurenche:1998gv}, nCTEQ15-np parton distribution functions corrected for nuclear matter effects \cite{Kovarik:2015cma}, and  BFG-II fragmentation functions \cite{Bourhis:1997yu}. The ``thermal photons'' are obtained by integrating 
photon emission rates \cite{Paquet:2015lta} over the entire spacetime volume occupied by the fluid-dynamical phase. Corrections to the photon emission rate from the effect of shear and bulk viscosities are available for many of the rates~\cite{Paquet:2015lta}, and are included when available.
Photon production in the hadronic cascade --- UrQMD --- has been studied in the past~\cite{Baeuchle:2009ep,Bauchle:2010ym} (see also Ref.~\cite{Huovinen:2002im}), although this feature is not included in its standard release.
As in several previous works, the photons from the late stages are computed here by using profiles from hydrodynamics down to T = 105 MeV.\footnote{Note that profiles from hydrodynamics in the low temperature regime 105 MeV $\leq$ T $\leq$ 145 MeV are only used for the calculations of photons. Hadrons are treated dynamically by UrQMD.} This procedure has been found to be justified quantitatively {\it a posteriori} by recent  microscopic calculations of photon production in the hadronic transport code SMASH \cite{Schafer:2019edr,*Schafer:2020vvw,*Schafer:2021slz}. 

We also need to describe photon production from the newly introduced \kompost phase.
As mentioned previously, the response functions  in the current implementation of \kompost were evaluated with pure-glue kinetic theory.
This could appear incomplete, since photon production requires electrically charged quarks to be present.
In practice, however, there is evidence that the response functions used in \kompost are similar across different microscopic theories~\cite{Giacalone:2019ldn,Kamata:2020mka}. 
In particular, the introduction of quarks does not appear to significantly change the response functions (Ref.~\cite{Giacalone:2019ldn} and references therein), and thus is not expected to have a major effect on the \emph{evolution} of the energy-momentum tensor in \kompostend.

In principle, computing photon production in the pre-hydrodynamics phase requires information beyond the energy-momentum tensor provided by \kompostend: one needs to know the photon emission rate in non-equilibrium nuclear matter.
In this work, we estimate photon emission in the \kompost stage in the same way as in the hydrodynamic stage: (i) the energy-momentum tensor is decomposed into energy density, flow velocity, bulk pressure and shear tensor; (ii) the temperature is calculated from the energy density with the QCD equation of state~\cite{Bazavov:2014pvz}; and (iii) photon emission rates are folded with the obtained temperature, flow velocity and viscous component profiles. While this approach has limitations, it ensures that photon emission is smooth at the transition to hydrodynamics. 

\begin{figure}[t!]
	\centering
	\includegraphics[width=0.85\linewidth]{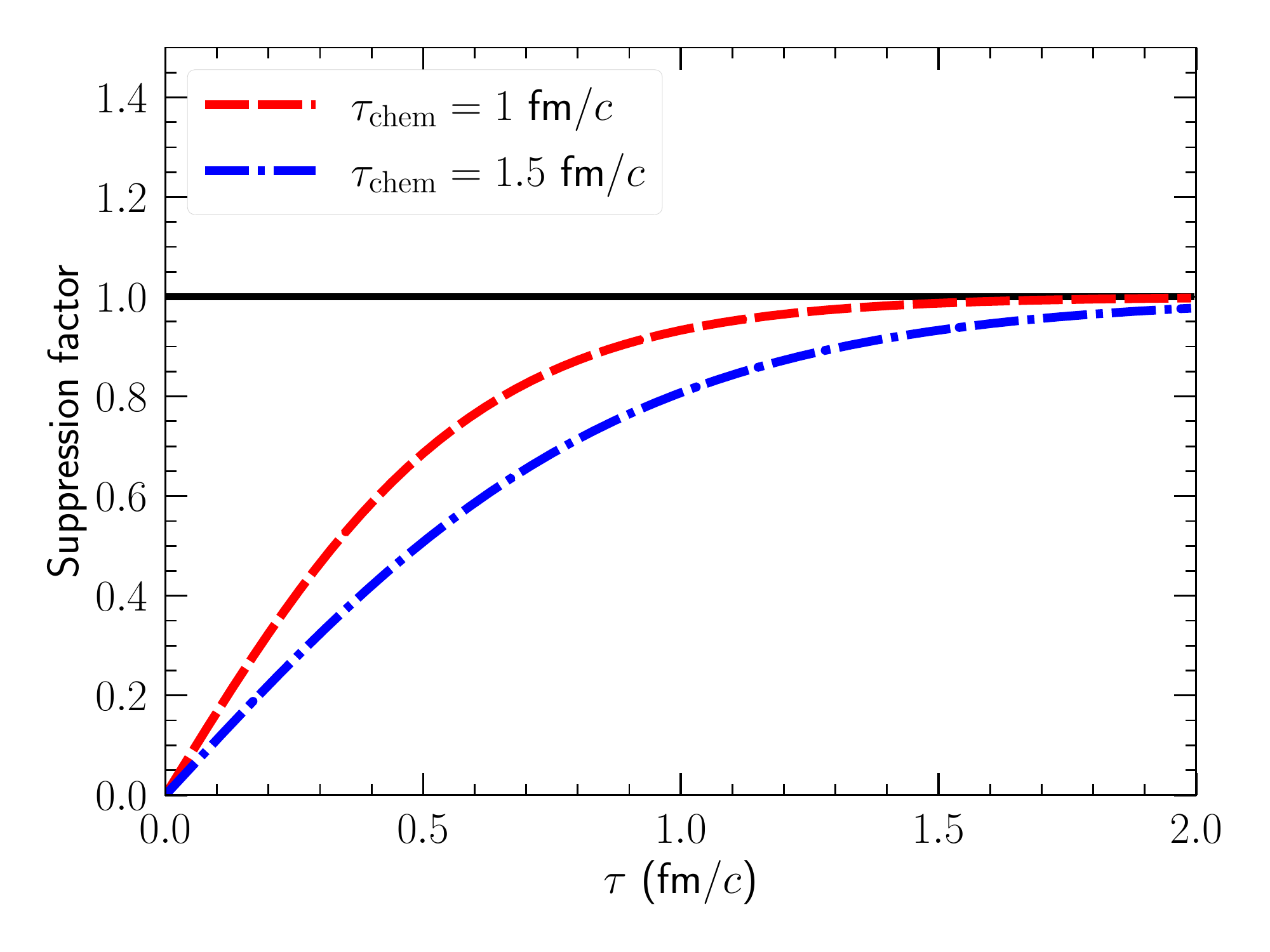}
	\caption{(Color online) The suppression factor applied to the thermal photon emission rates as a function of proper time, used to simulate the out-of-chemical equilibrium conditions. The two curves model the effect of a chemical equilibration time of 1 and 1.5 fm/$c$, respectively.
	}
	\label{fig:supress}
\end{figure}

The establishment of chemical equilibrium over time, i.e., how fast the quarks are produced and equilibrate in an initially purely gluonic system, can strongly affect the production of electromagnetic probes \cite{Traxler:1995kx,Chaudhuri_2000,Gelis:2004ep,Monnai:2014kqa,Bhattacharya:2015ada,Monnai:2015bca,Linnyk:2015rco,Greif:2016jeb,Vovchenko:2016ijt,Oliva:2017pri,Hauksson:2017udm,Monnai:2019vup,Churchill:2020uvk,Garcia_Montero_2020}.
QCD kinetic theory simulations~\cite{Kurkela:2018xxd,Kurkela:2018oqw} provide detailed information on the approach to chemical equilibrium. For the large coupling (low specific shear viscosity) such as those used in this work, Ref.~\cite{Kurkela:2018xxd,Kurkela:2018oqw} found that chemical equilibration is reached on a timescale slightly larger than the hydrodynamization time. Recall that our hydrodynamization time is the time we switch to hydrodynamics, namely $\tau_{\rm EKT}^{\rm hydro}$ or $\tau_{\rm IPG}^{\rm hydro}$. 

We account for  different possible chemical equilibration scenarios by multiplying the photon emission rate with a suppression factor at early times, based on the fermion energy density fraction of the equilibrium density from Ref.\,\cite{Kurkela:2018xxd}.
 We investigate three scenarios of photon production at early time: instant chemical equilibration, chemical equilibrium\footnote{Following Ref.~\cite{Kurkela:2018xxd}, chemical equilibrium is defined as the time at which the suppression factor reaches 0.9.} at $\tau_{\rm chem}=1$~fm/$c$,  or at $\tau_{\rm chem}=1.5$~fm/$c$. The corresponding  time-dependent suppression factors are shown in Fig.\,\ref{fig:supress}.

\begin{figure}[ht]
	\centering
	\includegraphics[width=0.85\linewidth]{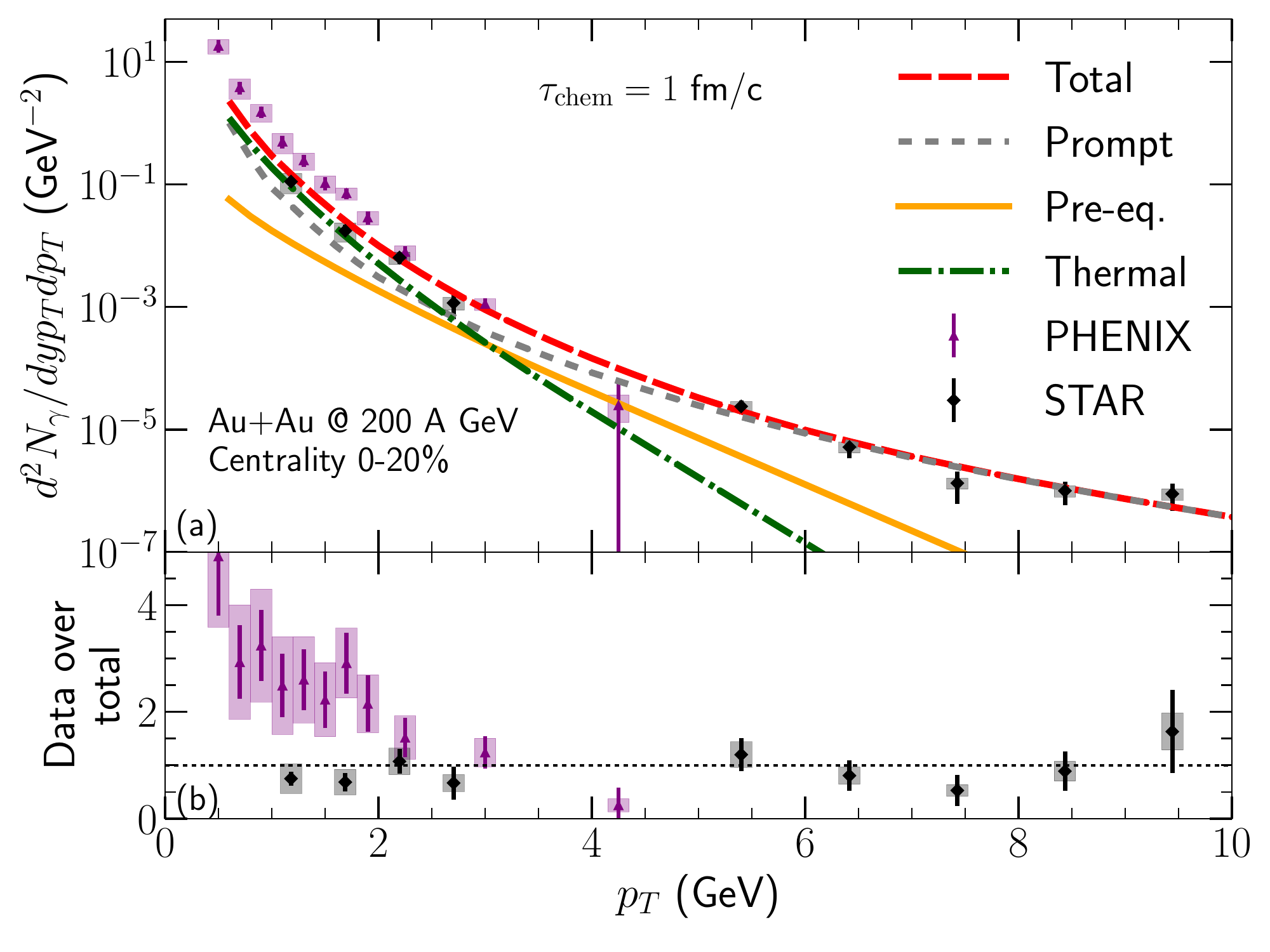}
	\caption{(Color online) 
	(a) The yield of direct photons in Au+Au collisions at maximum RHIC energy, in the $0 - 20\%$ centrality class. The different channels are enumerated in the text. Here, $\tau_{\rm chem} = 1$ fm/$c$. We compare to experimental data from the PHENIX \cite{Adare:2014fwh} and STAR \cite{STAR:2016use} Collaborations.
	(b) The ratio of experimental data over the total calculated photon yield.
	}
	\label{fig:rhic_photon_sources}
\end{figure}

The direct photon yield and the individual contributions from  different sources are shown in Figure~\ref{fig:rhic_photon_sources}. We show prompt photons (dashed curve), thermal photons (QGP + hadron phase, dashed-dotted curve), and photons from the pre-equilibrium stage (solid yellow curve). The total signal is the dashed red curve. The calculated photon yield is compared with measurements from the PHENIX \cite{Adare:2014fwh} and STAR \cite{STAR:2016use} collaborations. We note  the tension between the two experimental measurements for low momenta $(p_T \lesssim 2$ GeV), for the same system at the same centrality and energy.

Prompt photons that are produced earliest in the collision, dominate at high $p_T\gtrsim 3\,{\rm GeV}$; thermal photons are the largest contribution at $p_T\lesssim 3\,{\rm GeV}$. The pre-equilibrium contribution never dominates, but exceeds the thermal contribution for $p_T\gtrsim 3\,{\rm GeV}$.
The experimental data from the PHENIX collaboration is significantly underestimated at $p_T \lesssim 2$ GeV. A comparison to data from the STAR collaboration shows better agreement.


\begin{figure}[ht]
	\centering
	\includegraphics[width=0.85\linewidth]{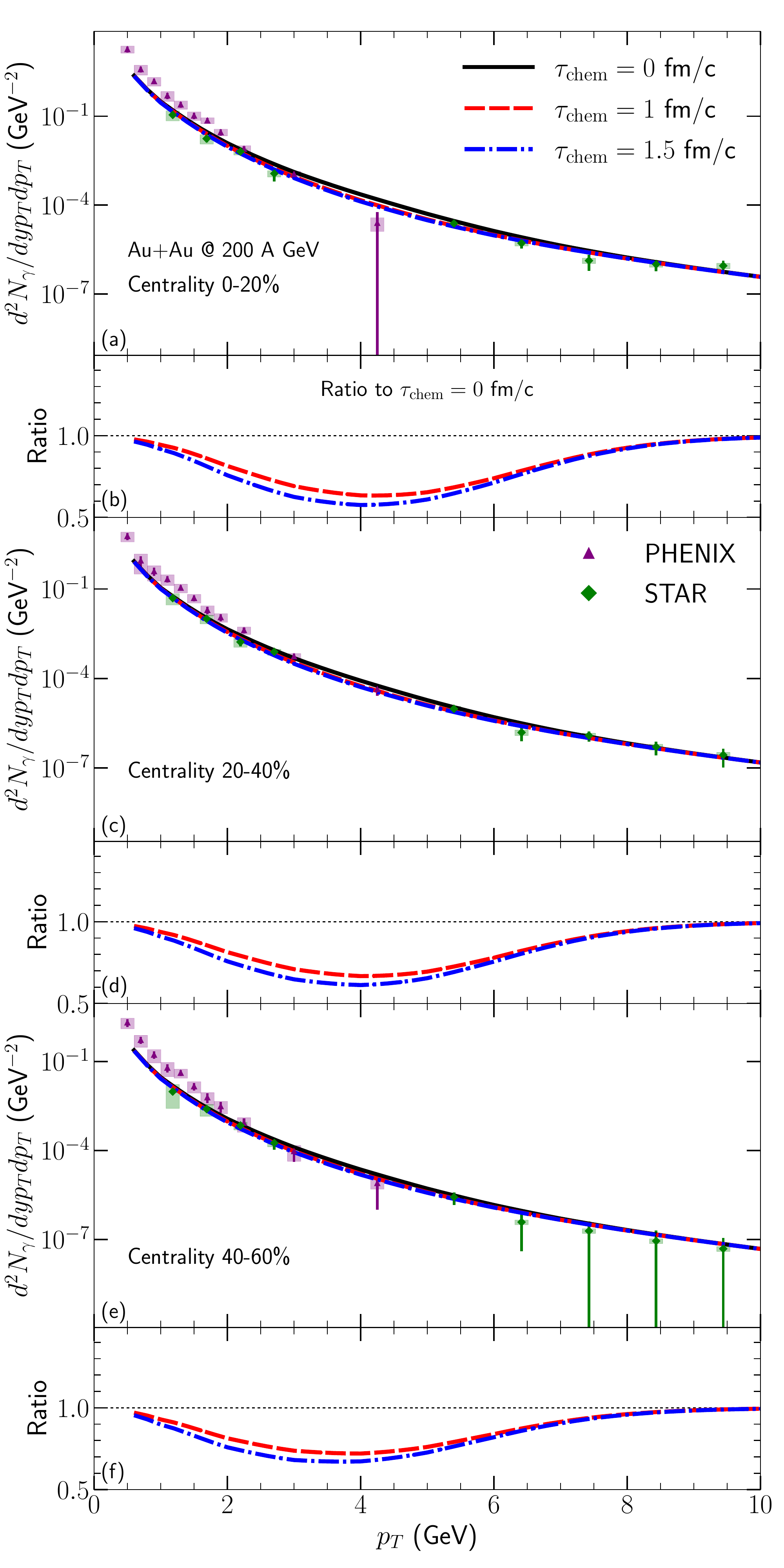}
	\caption{(Color online) 
	The photon yield for direct photons in Au+Au collisions at maximum RHIC energy, in different centrality classes, calculated for different values of the chemical equilibration time. The PHENIX data are from Ref.\,\cite{Adare:2014fwh} and the STAR data are from Ref.\,\cite{STAR:2016use}.}
	\label{fig:spectra_rhic_chem_photons}
\end{figure}

\begin{figure}[t]
	\centering
	\includegraphics[width=0.85\linewidth]{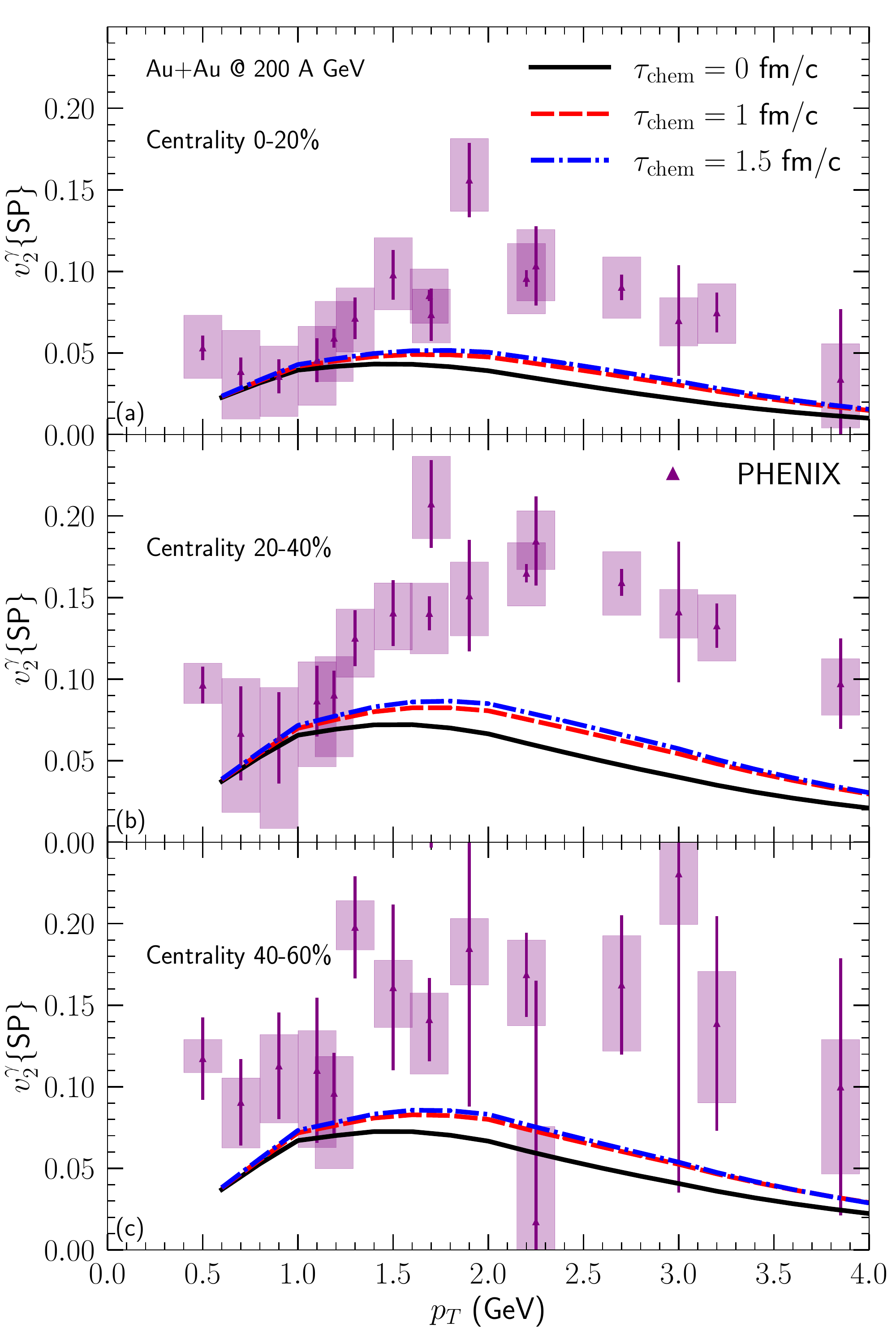}
	\caption{(Color online) 
	Same caption as for  Figure~\ref{fig:spectra_rhic_chem_photons}, but for the direct photon $v_2^\gamma\{SP\}$. Data are from Ref.\,\cite{Adare:2015lcd}.}
	\label{fig:v2_rhic_chem_photons}
\end{figure}

\begin{figure}[t]
	\centering
	\includegraphics[width=0.85\linewidth]{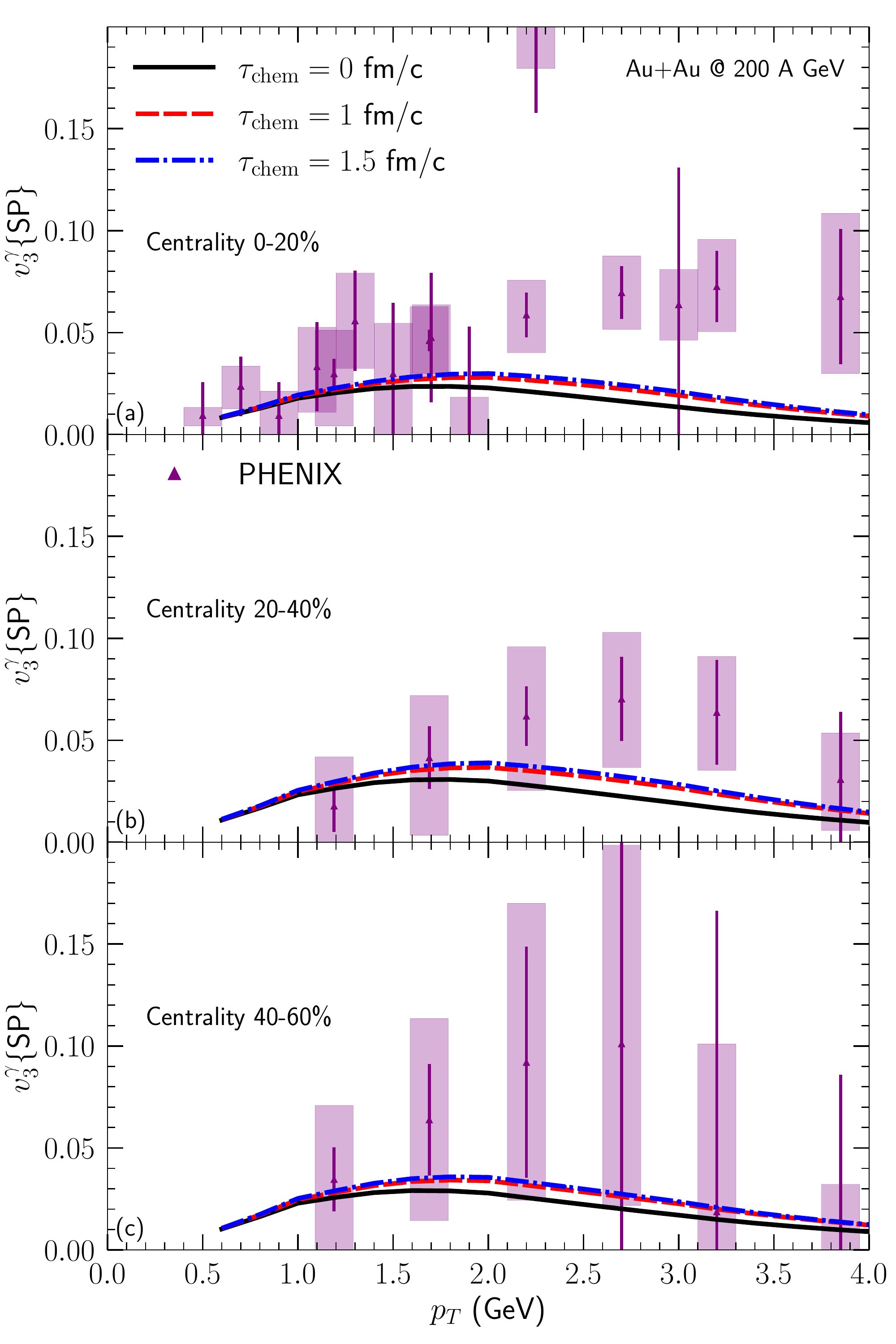}
	\caption{(Color online) 	Same caption as for  Figure~\ref{fig:spectra_rhic_chem_photons}, but for the direct photon $v_3^\gamma\{SP\}$. Data are from Ref.\,\cite{Adare:2015lcd}.}
	\label{fig:v3_rhic_chem_photons}
\end{figure}

The effect of the chemical non-equilibrium on the photon spectrum is shown in Fig.\,\ref{fig:spectra_rhic_chem_photons} for Au+Au collisions at RHIC energy. Panels (a),(c) and (e) correspond to the $0 - 20 \%$, $20-40\%$ and $40-60\%$ centrality class respectively.
These panels show the effect of suppressing the photon yield using the time-dependent factor shown in Fig.\,\ref{fig:supress} to reflect the dynamical buildup of the fermion content in the model.
The ratios to the instant chemical equilibration case are shown in panels (b), (d), and (f) for clarity.
The difference between a chemical equilibration time of 1 fm/$c$ or 1.5 fm/$c$ is small, and appears not to be resolved by the current RHIC data on photon spectra. The (chemical) non-equilibrium effects on the yield appear in the window $1\,{\rm GeV} \lesssim p_T \lesssim 7$ GeV for the physical conditions reported here. This window of intermediate $p_T$ is a result of later stage emission from the hadronic phase dominating at lower $p_T$, and of prompt photons dominating at higher $p_T$, neither of them being affected by the chemical equilibration.

Compared with experimental data from the PHENIX collaboration \cite{Adare:2014fwh}, the difference between calculation and measurement grows moving to more peripheral collisions, where it exceeds one standard deviation for $p_T \lesssim 2$~GeV in the $40 - 60\%$ centrality class. However, note that a comparison between the theoretical results reported here and the data reported by the STAR collaboration \cite{STAR:2016use}, for the same system at the same energy, show better agreement, as also shown previously in Fig.\,\ref{fig:rhic_photon_sources}. Theory and STAR measurements agree within uncertainties, for the two centralities addressed by the experimental collaboration. The origin of the discrepancy between the photon results reported by STAR and PHENIX at RHIC remains unresolved to this day \cite{David:2019wpt}.

The corresponding calculations for the photon elliptic and triangular momentum anisotropies,  $v_2^\gamma$ and  $v_3^\gamma$, are shown in Figures~\ref{fig:v2_rhic_chem_photons} and \ref{fig:v3_rhic_chem_photons}, respectively, and compared to experimental data from the PHENIX collaboration.\footnote{Note that the STAR collaboration has not released measurements of photon $v_n$.} Importantly, photon momentum anisotropy coefficients are not obtained through photon-photon correlations but rather using photon-hadron correlations, because of statistics considerations. The analysis techniques differ between PHENIX and ALICE (to be shown shortly): the former uses the event-plane method, while the latter uses the scalar product method.
As discussed in detail in Ref.~\cite{Paquet:2015lta}, we calculate the photon $v_2^\gamma$ with the scalar product method to compare with both PHENIX and ALICE:
\bea
v_n^\gamma\left\{ SP \right\} \left(p_T^\gamma\right) = \frac{\langle v_n^\gamma \left( p_T^\gamma\right) v_n^h \cos \left( n \left( \Psi_n^\gamma (p_T^\gamma) - \Psi_n^h\right)\right) \rangle}{\sqrt{ \langle \left(v_n^h\right)^2\rangle}}
\eea
where $h$ refers to hadrons. The measured event-plane flow anisotropy is close to the scalar product results in the low resolution limit \cite{Luzum:2012da}.

The effect of delayed chemical equilibration on the photon spectra naturally leads to a suppression of the number of early-stage photons as shown in Figure~\ref{fig:spectra_rhic_chem_photons}. For the $v_2^\gamma$ and $v_3^\gamma$ of photons, Figures~\ref{fig:v2_rhic_chem_photons} and \ref{fig:v3_rhic_chem_photons} show that a delayed equilibration time increases both $v_{2/3}^\gamma$ in the $1-4$~GeV $p_T$ range. This is a straightforward consequence of suppressing early-stage photons that have a small $v_{2/3}^\gamma$ and would dilute the larger $v_{2/3}^\gamma$ of photons emitted at later times. As such, larger values of chemical equilibration time do improve slightly the agreement with experimental measurements.


\begin{figure}[t]
	\centering
	\includegraphics[width=0.85\linewidth]{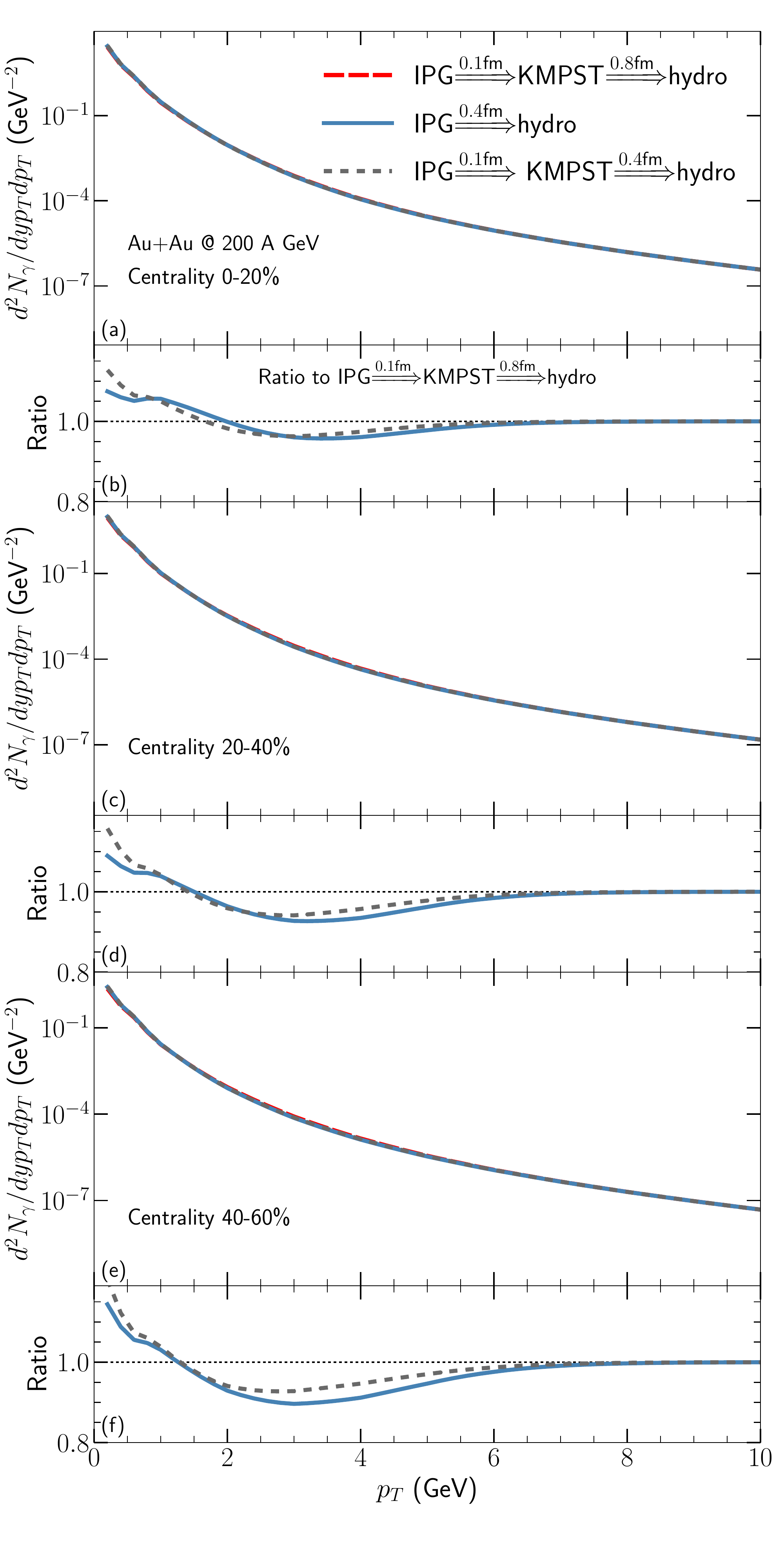}
	\caption{(Color online) The photon yield for direct photons in Au+Au collisions at maximum RHIC energy, in three different centrality classes. Here, $\tau_{\rm chem} = 1$ fm/$c$. See main text for details. }
	\label{fig:spectra_photon_init_scenarios}
\end{figure}		

\begin{figure}[t]
	\centering
	\includegraphics[width=0.85\linewidth]{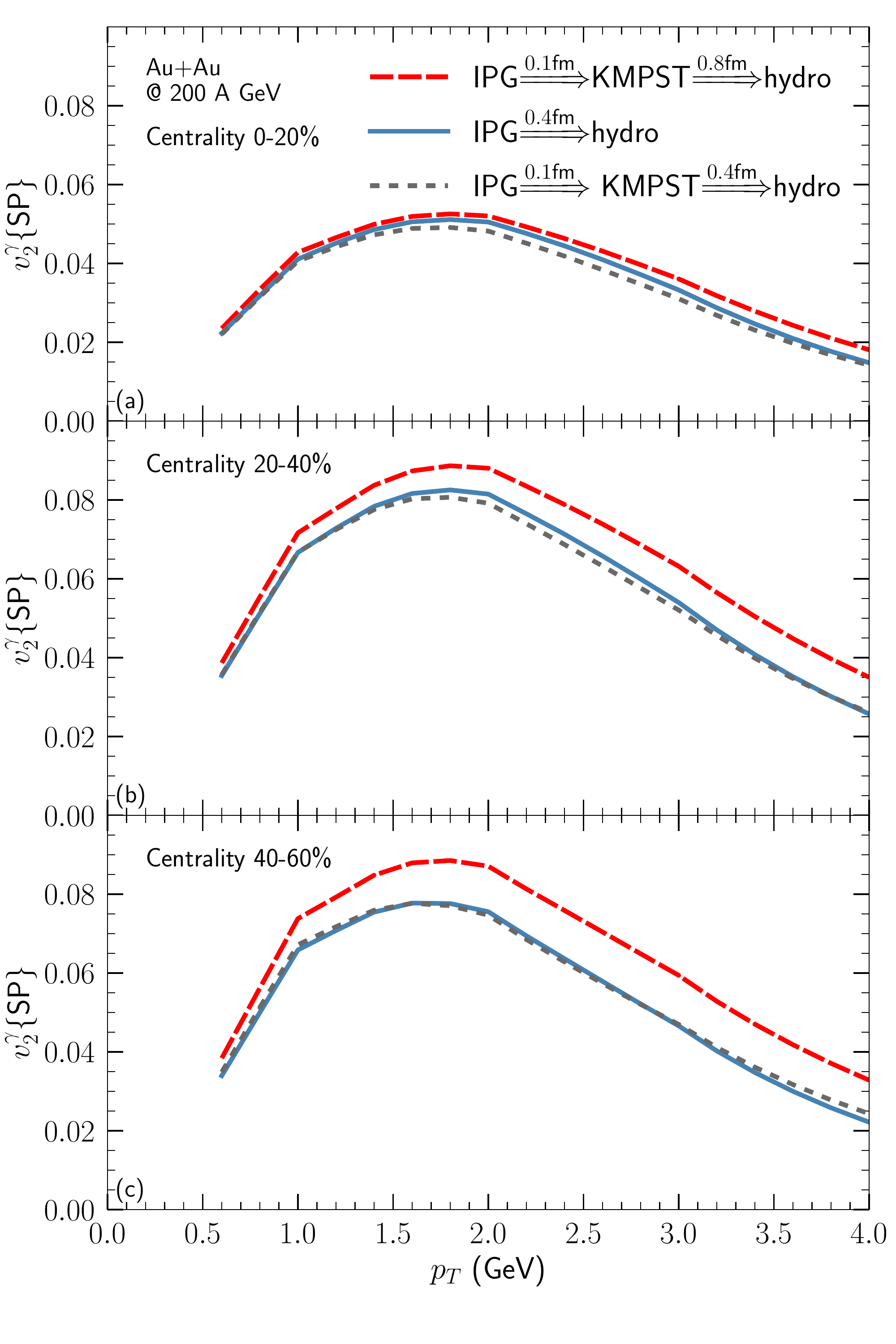}
	\caption{(Color online) Same caption as for Fig.\,\ref{fig:spectra_photon_init_scenarios}, but for the photon elliptic flow $v_2^\gamma (p_T)$. }
	\label{fig:v2_photon_init_scenarios}
\end{figure}

The effect of the \kompost phase in the photon spectrum is highlighted in Fig.\,\ref{fig:spectra_photon_init_scenarios}. Starting from the top of the legend, the first curve shows photons generated in our scenario I, with $\tau_{\rm IPG}^{\rm EKT}=0.1$~fm/$c$ and $\tau_{\rm EKT}^{\rm hydro}=0.8$~fm/$c$. The second curve shows the direct photon signal obtained in scenario II with $\tau_{\rm IPG}^{\rm hydro}=0.4$~fm/$c$. Finally, in order to obtain a more meaningful assessment of the role of the \kompost phase with respect to scenario II, a calculation run with the pre-equilibrium  dynamics lasting only up to 0.4 fm/$c$ is done ($\tau_{\rm IPG}^{\rm EKT}=0.1$~fm/$c$ and $\tau_{\rm EKT}^{\rm hydro}=0.4$~fm/$c$). The ratios of both this last scenario and scenario II to scenario I are also shown in Fig.\,\ref{fig:spectra_photon_init_scenarios}. Here it is revealed that differences are maximally 10\%, with the spectra enhanced by a longer \kompost phase for $1.5\,{\rm GeV} < p_T < 6\,{\rm GeV}$ and reduced for $p_T\lesssim 1.5\,{\rm GeV}$. Replacing IP-Glasma with \kompost between 0.1\,fm/$c$ and 0.4\,fm/$c$ has a small effect.

The corresponding result for the photon $v_2^\gamma$ is shown in Figure~\ref{fig:v2_photon_init_scenarios}, for 0-20\%, 20-40\% and 40-60\% centralities.
For the most central case, all three pre-equilibrium scenarios lead to similar $v_2^\gamma$. However, a different result is observed for more peripheral centralities: the $v_2^\gamma$ when transitioning from \kompost{} to hydrodynamics at $\tau_{\rm EKT}^{\rm hydro}=0.8$~fm/$c$ is larger than the two other scenarios that transition to hydrodynamics at $0.4$~fm/$c$.
This larger dependence on the pre-equilibrium scenarios in peripheral collisions was also observed for hadrons. In general, peripheral collisions are expected to be more sensitive to the pre-equilibrium phase, as a consequence of their shorter hydrodynamic phase compared to central events.
Physically, the dependence of $v_2^\gamma$ on the transition time to hydrodynamics could be related to the conformal nature of the pre-equilibrium models. 

The overall difference between the three scenarios is small compared to the current tension with PHENIX measurements.
In this sense, current RHIC photon data does not demand the presence of a phase like K\o{}MP\o{}ST.
However, and importantly, the \kompost phase has the potential to provide a dovetail matching \cite{Kurkela:2018xxd} between IP-Glasma and relativistic viscous fluid dynamics, from a theory point of view.
Including a pre-equilibrium phase can also avoid non-causal behavior, which can appear when running hydrodynamics using a general off-equilibrium initial state like IP-Glasma \cite{Cheng:2021tnq,Plumberg:2021bme}.
The matching to hydrodynamics is expected to be further improved by future non-equilibrium models that go beyond current approximations such as conformality and linear response.

\begin{figure}[t]
	\centering
	\includegraphics[width=0.95\linewidth]{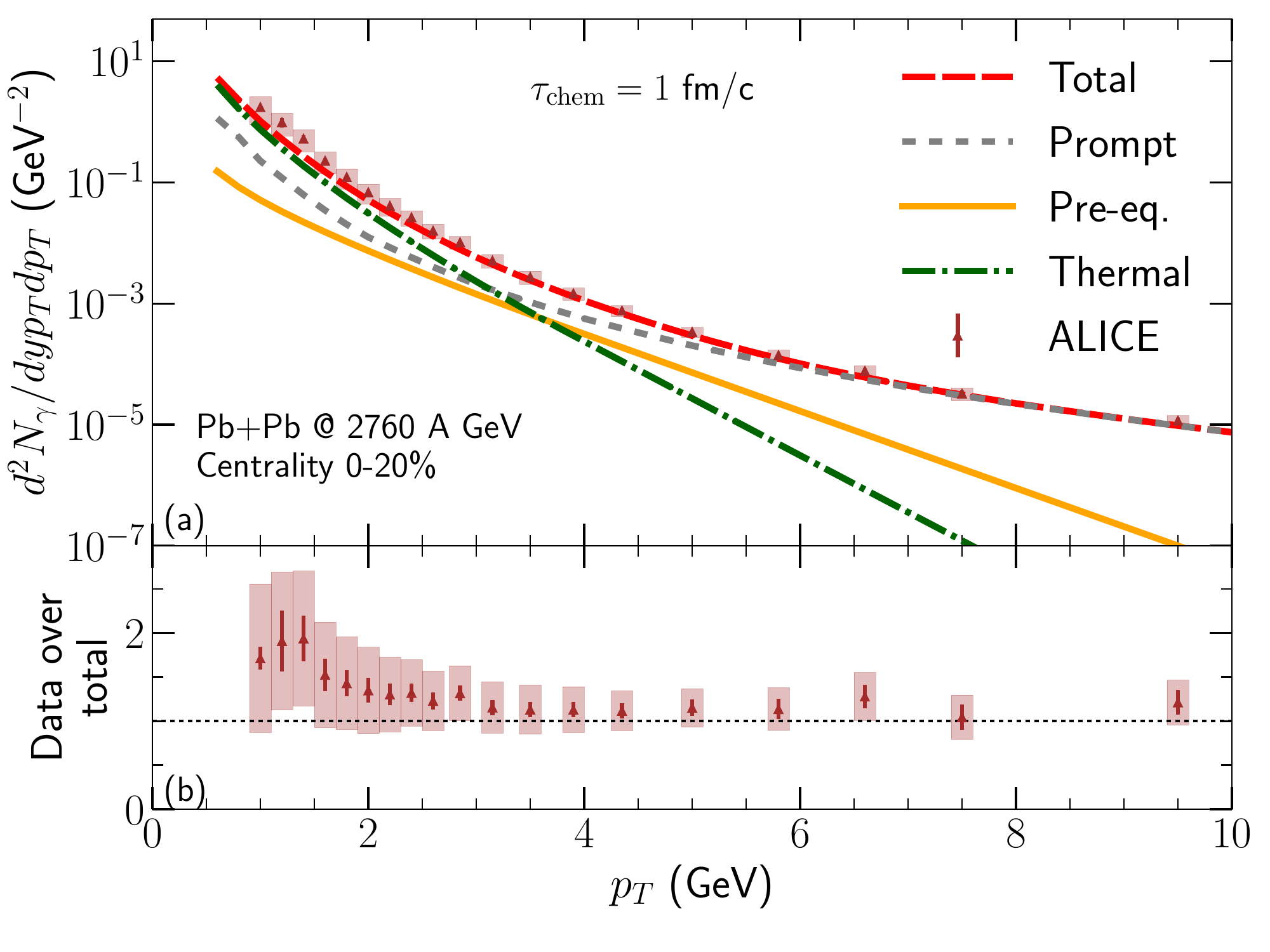}
	\caption{(Color online) (a) Direct photon yield in Pb+Pb collisions at 2760 A GeV, in the $0 - 20\%$ centrality class. Also shown is the breakdown into the different components. (b) The ratio between the experimental data and the calculated total photon yield. Experimental data from the ALICE Collaboration \cite{Adam:2015lda}. See main text for details. }
	\label{fig:LHCphoton_no}
\end{figure}
Considering photons generated by collisions at the LHC, Fig.\,\ref{fig:LHCphoton_no} (a) reports on results of calculations performed for Pb+Pb at 2760 A GeV. As in the case of RHIC, the same hydrodynamic evolution is used for photons as for hadrons. The kinematical window where the pre-equilibrium photons contribute the most (relative to prompt and thermal photon yields) is similar to that seen at RHIC, around $p_T = 4$ GeV. 
Experimental data from the ALICE Collaboration \cite{Adam:2015lda} is shown for comparison. Figure \ref{fig:LHCphoton_no} (b) shows the ratio of the experimental data to the calculated total direct photon yield. Agreement is very good for $p_T\gtrsim 3\,{\rm GeV}$, with deviations, albeit mostly within experimental errors, seen at lower $p_T$.

\begin{figure}[t]
	\centering
	\includegraphics[width=0.9\linewidth]{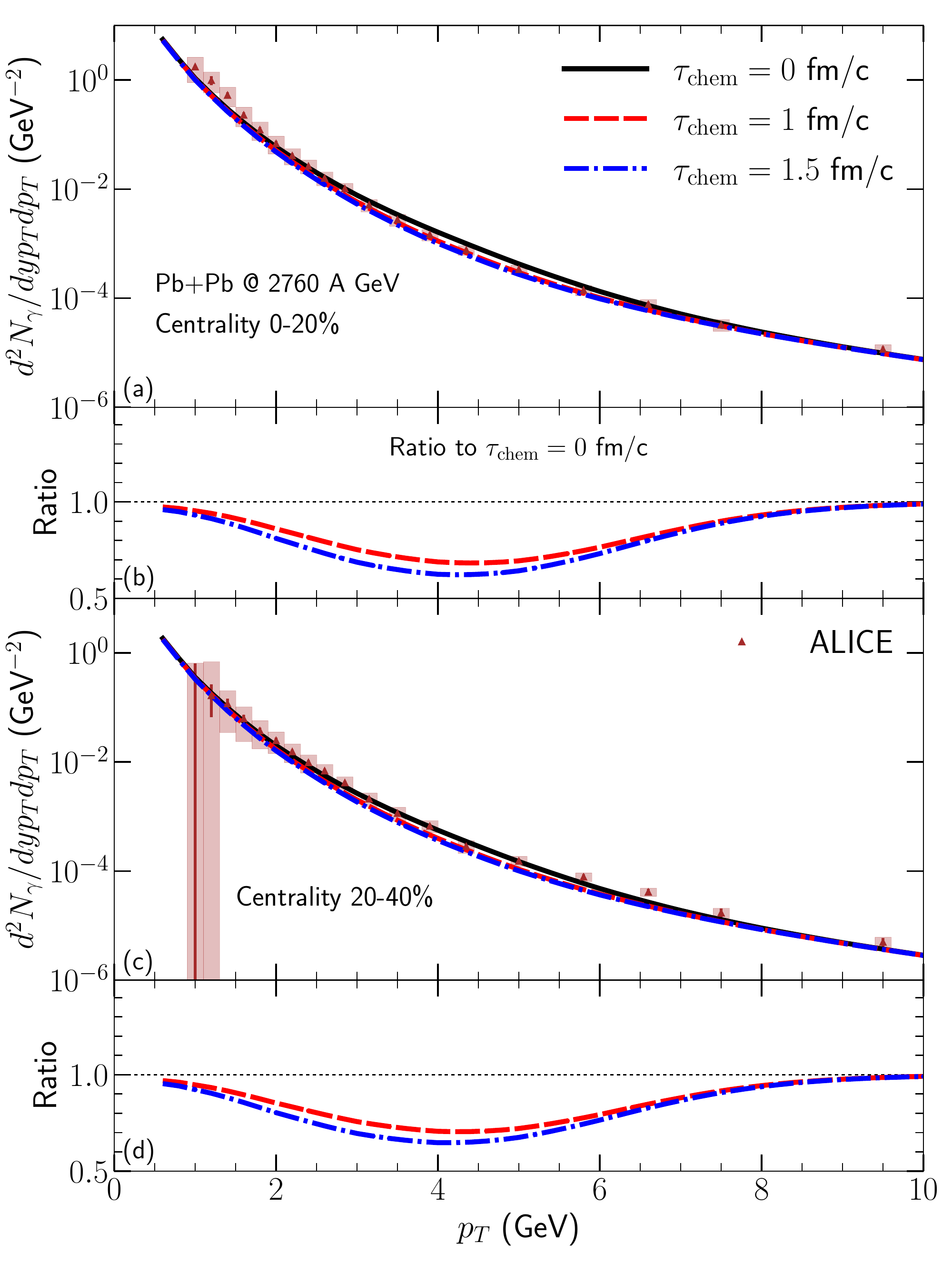}
	\caption{(Color online) The photon yield for direct photons in Pb+Pb collisions at $\sqrt{s_\mathrm{NN}} = 2760$\,GeV in three different centrality classes. Here, $\tau_{\rm chem} = 1$ fm/$c$. See main text for details. }
	\label{fig:LHCKomPhot}
\end{figure}

The effect of chemical non-equilibrium on the direct photon yields in two different centrality classes is illustrated in Fig.\,\ref{fig:LHCKomPhot}. 
The chemical suppression factor has the largest effects in the window $2 \lesssim p_T \lesssim 7$ GeV, and the ALICE data show a  preference for the model with chemical non-equilibrium. 

\begin{figure}[t]
	\centering
	\includegraphics[width=0.9\linewidth]{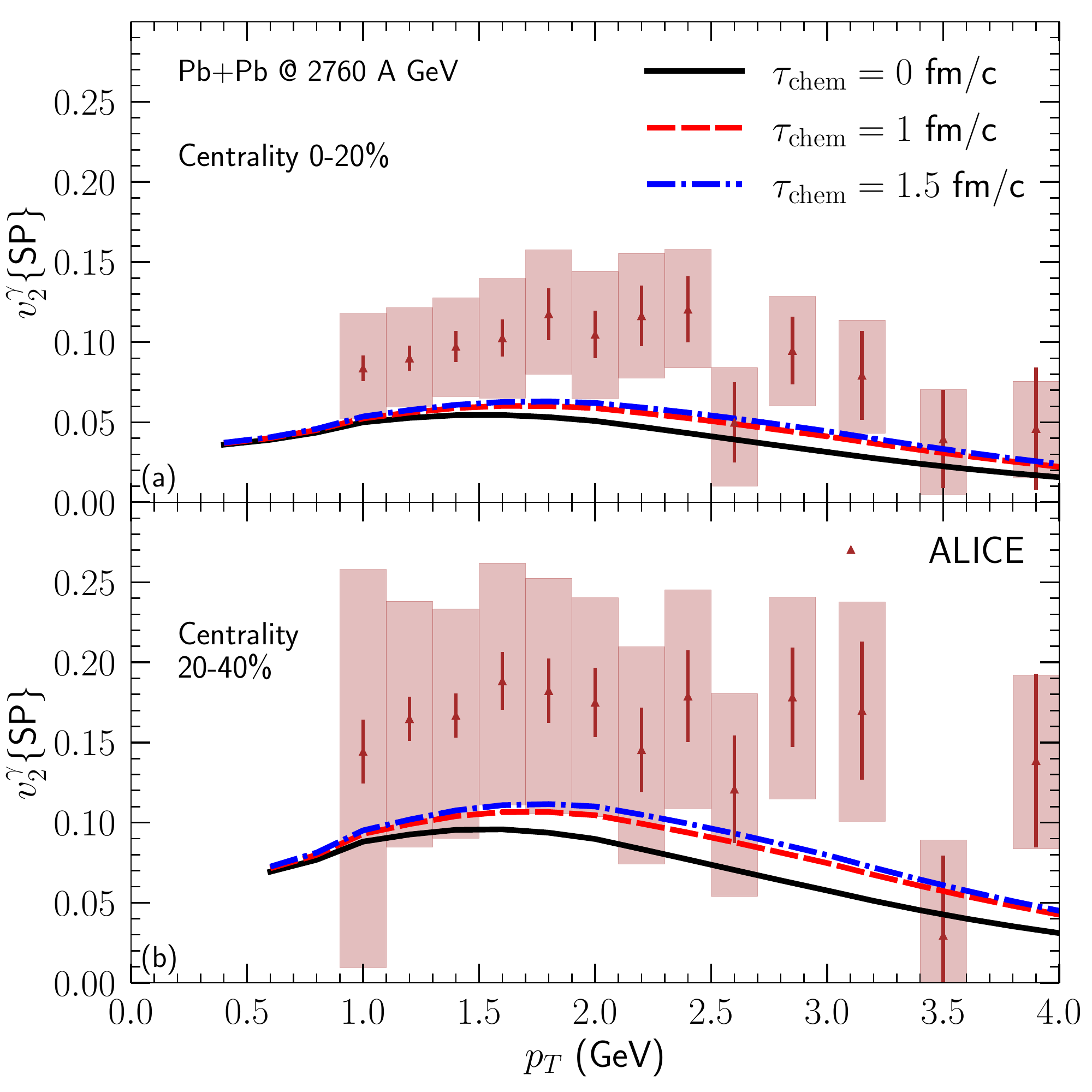}
	\caption{(Color online) Direct photon elliptic flow $v_2^\gamma (p_T)$ in $2760\,{\rm A GeV}$ Pb+Pb collisions for different chemical equilibration times compared to experimental data from the ALICE collaboration \cite{Acharya:2018bdy}. See main text for details. }
	\label{fig:LHCphoton_v2}
\end{figure}

Figure \ref{fig:LHCphoton_v2} shows the elliptic flow of direct photons in 2760\,{\rm A GeV} Pb+Pb collisions, compared to experimental data from the ALICE collaboration \cite{Acharya:2018bdy}. For the measured centrality classes, the calculations systematically undershoot the data, even if the two are mostly consistent within uncertainties. 
As was the case at RHIC, a larger value of the chemical equilibration time reduces the difference between theory and experimental data.

\begin{figure}[t]
	\centering
	\includegraphics[width=0.85\linewidth]{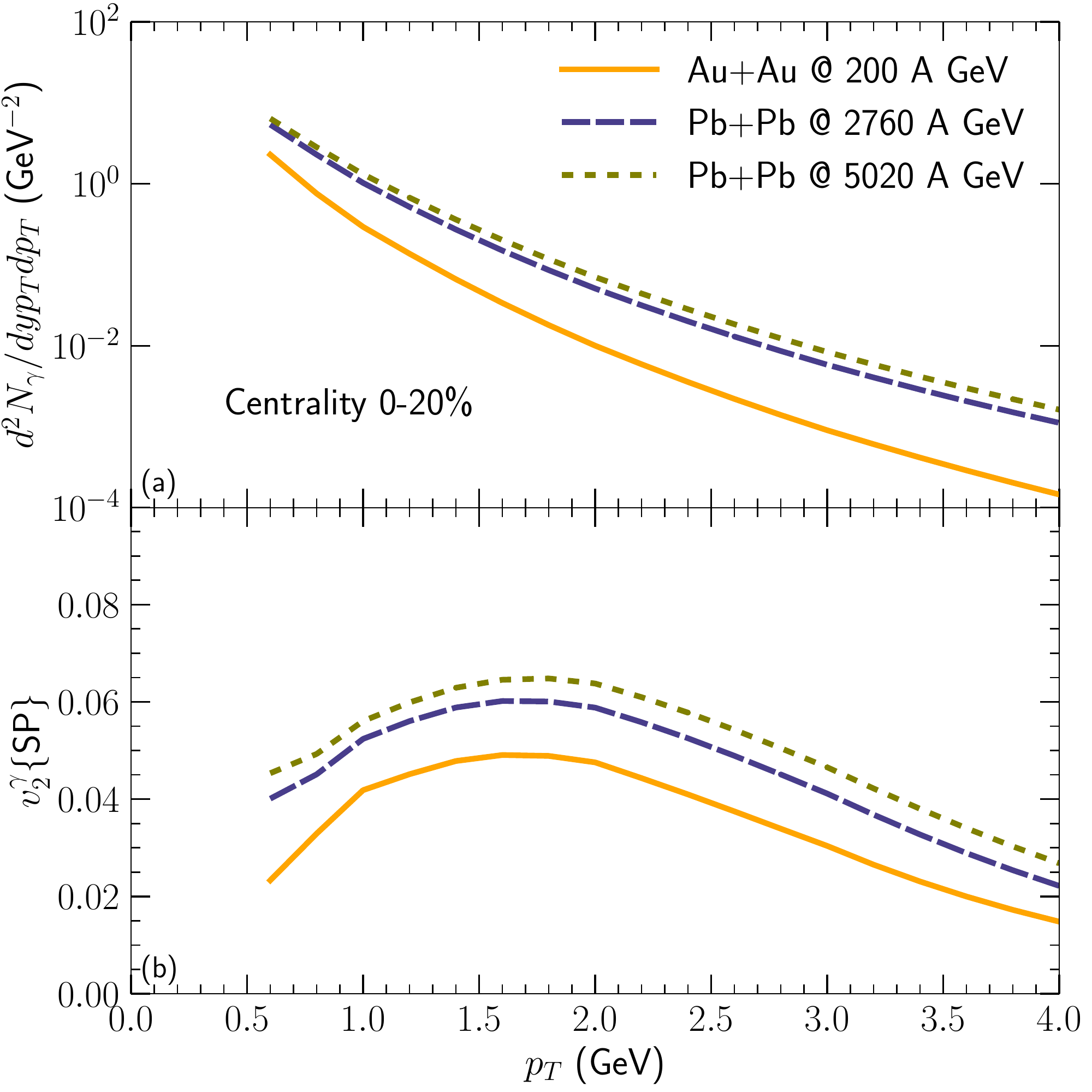}
	\caption{(Color online) The direct photon yield and $v_2^\gamma$ in 0-20\% centrality, for Au+Au @ 200 A GeV and Pb+Pb @ 2760 A GeV and 5020 A GeV.}
	\label{fig:LHCKomPhot_all}
\end{figure}

Figure~\ref{fig:LHCKomPhot_all} shows the collision energy dependence of the direct photon spectra and their elliptic flow coefficients from the top RHIC to LHC energies in 0-20\% central heavy-ion collisions. We find that direct photon yield and $v_2^\gamma$ increase with $\sqrts$. Prompt photons have flatter $p_T$-spectra at LHC energies. At the same time, more thermal photons radiate from Pb+Pb collisions at the LHC than those at RHIC because of the higher averaged fireball temperature and larger space-time volume. The increase of direct photon $v_2^\gamma$ with increasing collision energy suggests that the enhancement in the thermal radiation wins over the prompt photon production at high energies.

\subsection{Photon production in ``small systems''}

This section is devoted to the theoretical treatment of what has become known in the field as ``small systems''.
This includes asymmetric systems involving protons and a heavier ion, as well as symmetric collisions of two smaller nuclei, such as O+O. The interest generated by such configurations owes much to the fact that those systems show a similar degree of collectivity~ ({\it e.g.} flow coefficients) than that observed in ``large'' colliding systems such as some of the ones discussed earlier: Au+Au and Pb+Pb. This fact was met with surprise by the community, as one of the original motivations to perform experimental runs with proton projectiles incident on a larger target was to provide a neutral baseline. 
In addition to their undeniable theoretical interest, small systems deserve a separate section here because \kompost was not designed to operate in such environments: the combination of small size and large gradients pushes the linear treatment past its limit of validity. Therefore, only the combination of IP-Glasma and hydrodynamics --- scenario II --- is used in this section. 

\begin{figure}[ht!]
	\centering
	\includegraphics[width=0.9\linewidth]{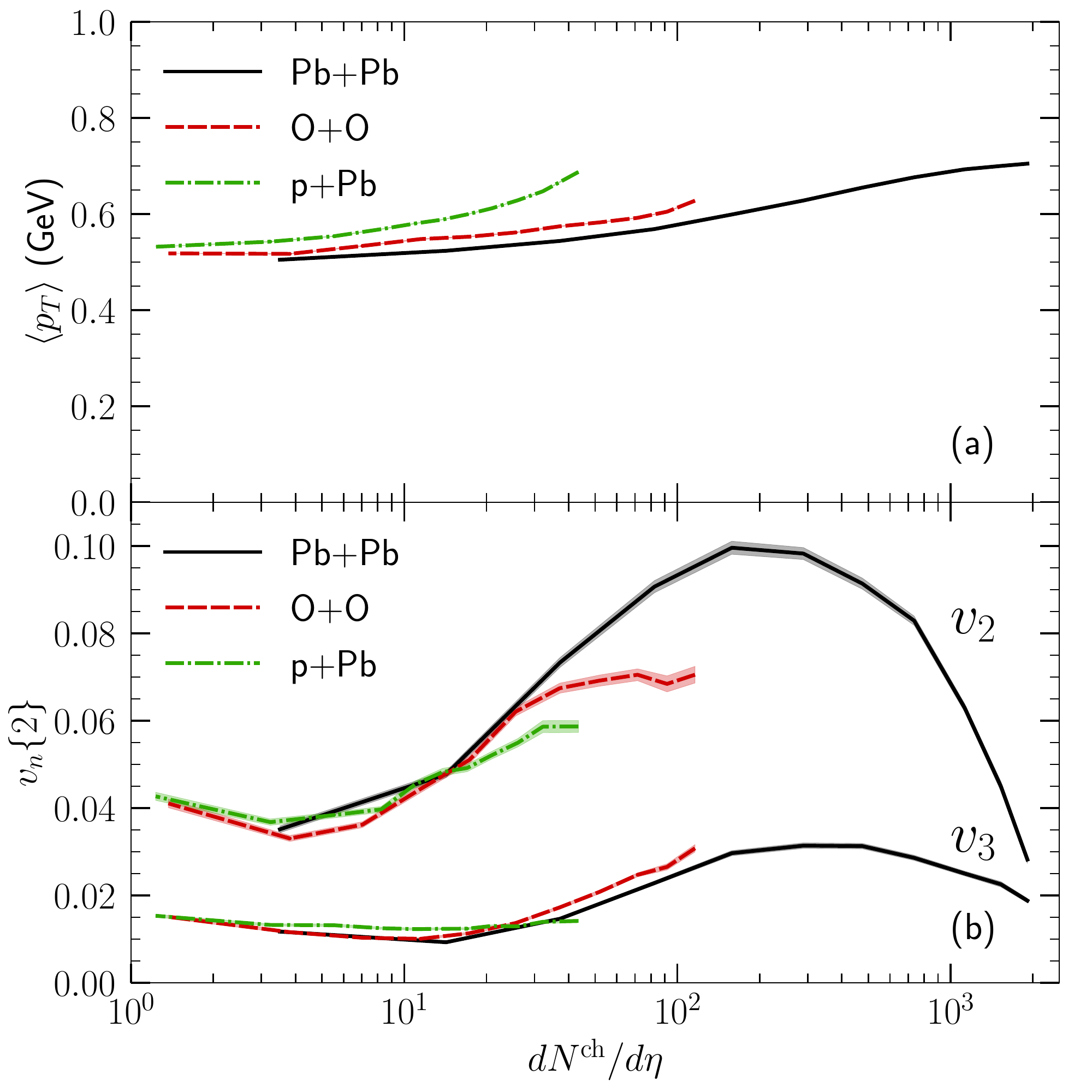}
	\caption{(Color online) Averaged transverse momentum and flow anisotropic coefficients for charged hadrons in O+O and p+Pb collisions compared with those in Pb+Pb collisions at 5020 A GeV.}
	\label{fig:SmallSystemFlow}
\end{figure}

In addition to investigating whether QGP is also present in small systems, another reason to study them is that they provide another mean to explore the mechanisms of particle production as a function of system size. For instance, for a colliding energy of 5020 A GeV, the charged hadron multiplicity, $d N_{\rm ch}/d \eta$,for minimum-bias O+O collisions is comparable to that for Pb+Pb collisions in a centrality class of $70-80\%$ \cite{Huss:2020whe, Brewer:2021kiv}. Similarly, O+O collisions at the same energy and in a centrality class of $0 -5\%$ can be compared to Pb+Pb collisions in a $50-70\%$ centrality class.

The calculated charged multiplicity dependence of light system (O+O and p+Pb) observables $\left(\langle p_T \rangle, v_n \{ 2 \}, n=2,3\right)$ is shown in Fig.\,\ref{fig:SmallSystemFlow}. Hadrons' averaged transverse momenta in p+Pb and O+O collisions are larger than those in Pb+Pb collisions at the same multiplicity, which indicates that a stronger radial flow is developed by larger pressure gradients in the more compact systems. For $dN^\mathrm{ch}/d\eta > 20$, the elliptic flow coefficients have the ordering $v_2(\mathrm{Pb+Pb}) > v_2(\mathrm{O+O}) > v_2(\mathrm{p+Pb})$, driven by the global collision geometry. In contrast, the triangular flow $v_3$ coefficients do not show any system dependence at fixed charged particle multiplicities since $v_3$ is mainly driven by the event-by-event initial-state fluctuations. We note that both $v_2$ and $v_3$ increase as $dN^\mathrm{ch}/d\eta$ goes below $5$. In these peripheral collisions, the initial-state momentum anisotropy gradually begins to dominate the measured anisotropic flow coefficients over the hydrodynamic response to the global geometry \cite{Schenke:2020mbo,Giacalone:2020byk}.

\begin{figure}[ht!]
	\centering
	\includegraphics[width=0.9\linewidth]{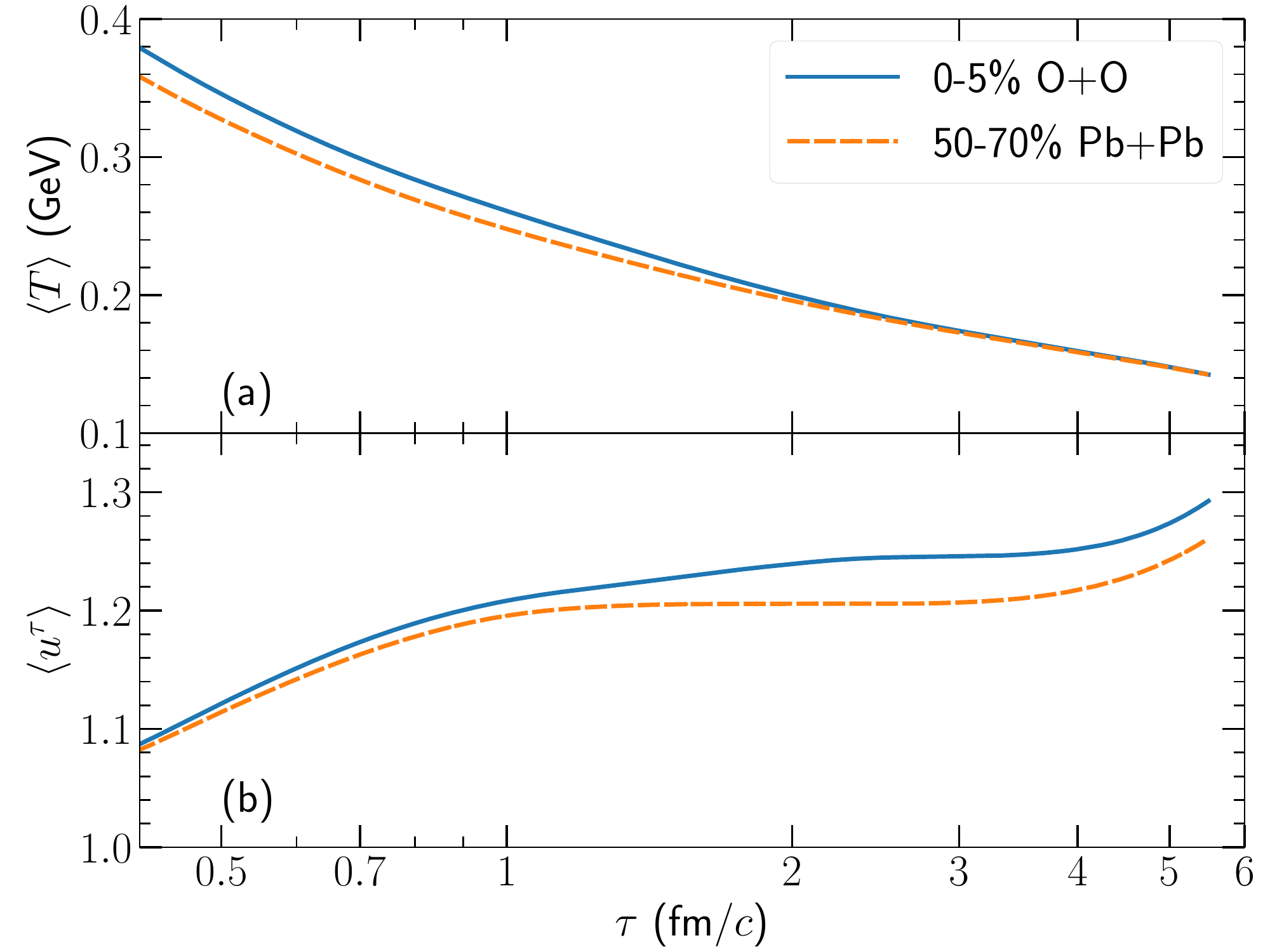}
	\caption{(Color online) (a) Energy-density-weighted average temperature of fluid dynamical simulations for central O+O and semi-peripheral Pb+Pb collisions with the same charged hadron multiplicity at 5020 A GeV. (b) Energy density weighted average transverse flow velocity for the two systems described in (a).}
	\label{fig:T_OO}
\end{figure}
Figure~\ref{fig:T_OO} shows the evolution of energy-density-weighted average temperature (a) and flow velocity (b) in 0-5\% O+O and 50-70\% Pb+Pb collisions. As already mentioned, these two collision systems produce similar charged hadron multiplicity at midrapidity, $dN^\mathrm{ch}/d\eta = 116$. Owing to a relatively  compact size, central O+O collisions achieve a higher average temperature and stronger radial flow during the hydrodynamic evolution than those in the semi-peripheral Pb+Pb collisions. The time-integrated effect of such a stronger radial flow is reflected in the hadrons' mean transverse momenta in Fig.\,\ref{fig:SmallSystemFlow} (a). However, it is difficult to find experimental evidence of the higher initial temperature in the central O+O collisions using hadronic observables. In this case, thermal photon radiation provides a better tool to access the early-stage dynamics.

\begin{figure}[t]
	\centering
	\includegraphics[width=0.85\linewidth]{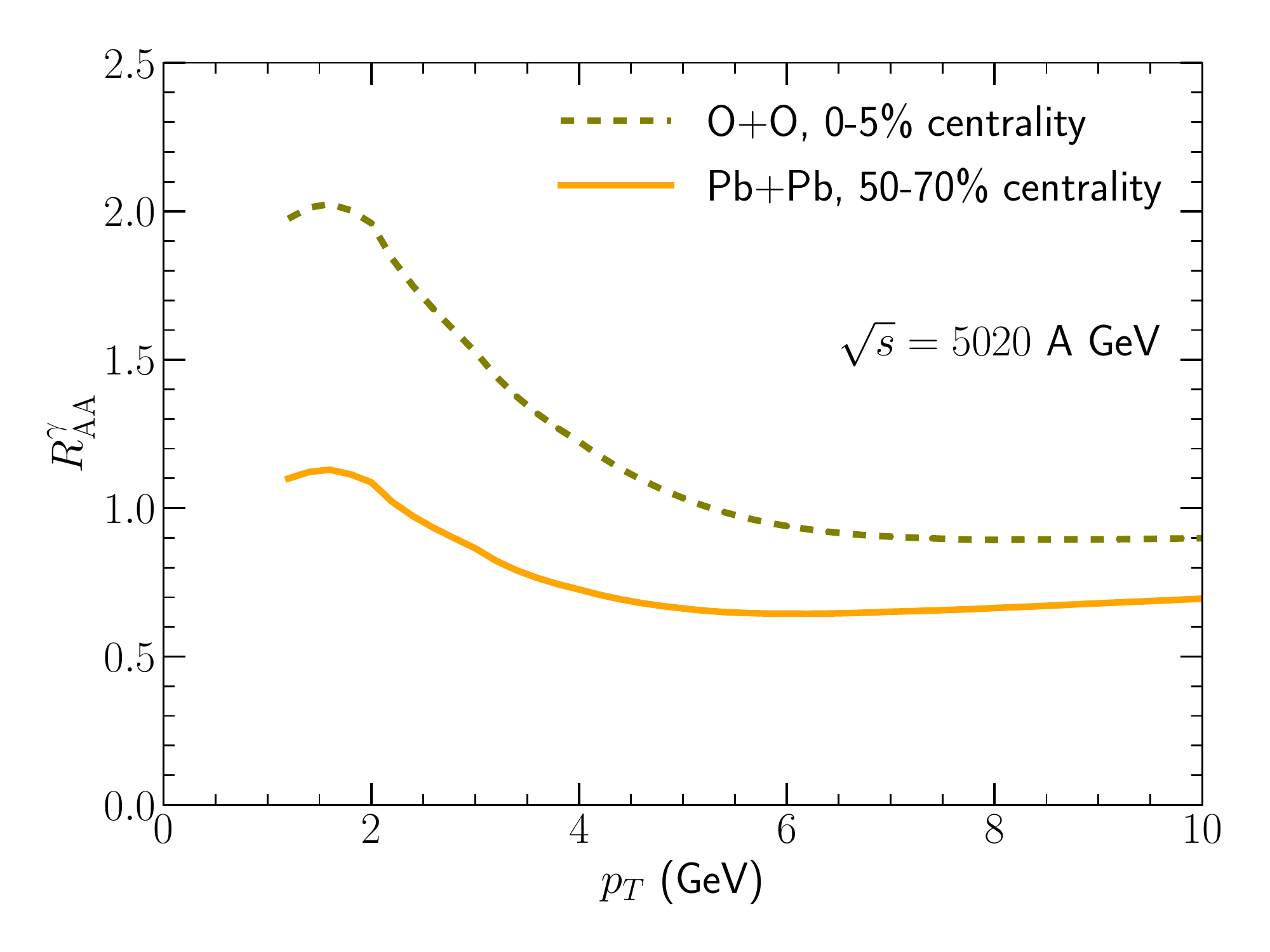}
	\caption{(Color online) $R_{AA}^\gamma$ of photon in 0-5\% O+O collisions at 5020 A GeV, compared to 50-70\% Pb+Pb collisions.}
	\label{fig:RAA_OO_PbPb_comp}
\end{figure}

For the same conditions -- central O+O collisions and peripheral Pb+Pb collisions -- the direct photon nuclear modification factor, $R^\gamma_{AA}$ is shown in Fig.\,\ref{fig:RAA_OO_PbPb_comp}. This figure illustrates well the power of electromagnetic radiation to report on local conditions at the time of its emission. The difference between  central O+O and peripheral Pb+Pb is now clearly visible when comparing the photon nuclear modification ratios. For $p_T \sim 2$~GeV, $R_{AA}^\gamma$ for  O+O collisions in the $0-5 \%$ centrality class  is enhanced by approximately 80\%  over its value in peripheral Pb+Pb events, for $\sqrt{s_{\rm NN}}$ = 5020 A GeV.  

\begin{figure}[t]
	\centering
	\includegraphics[width=0.85\linewidth]{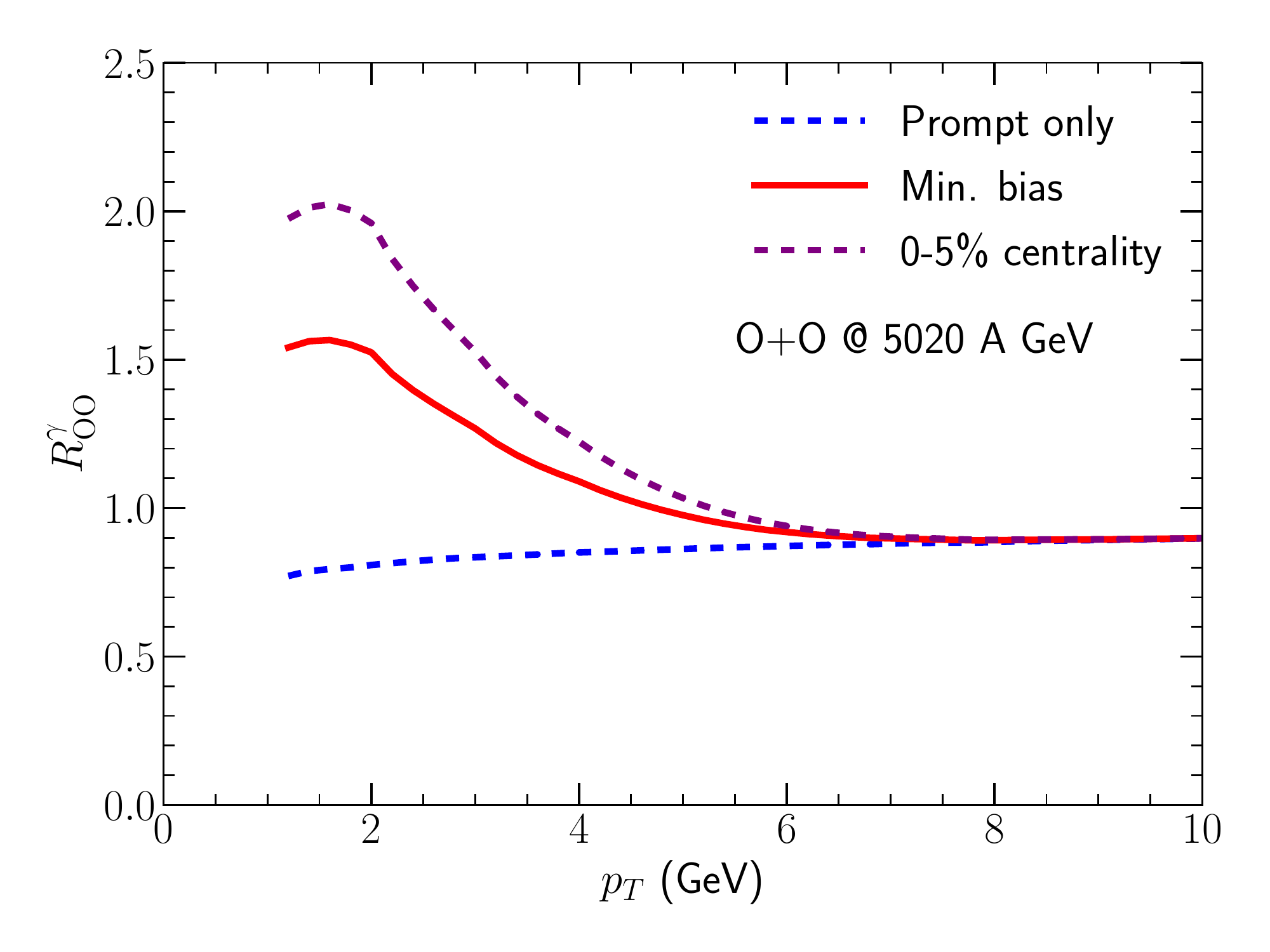}
	\caption{(Color online) The $R^\gamma_{AA}$ of direct photons in O+O collisions at 5020 A GeV. The lower line includes only prompt photons, with isospin effect and nCTEQ15 nuclear pdf's as described in Section~\ref{results:photons}. The middle and top lines include medium photons, for minimum bias collisions and 0-5\% central collisions, respectively.}
	\label{fig:RAA_OO}
\end{figure}

\begin{figure}[t]
	\centering
	\includegraphics[width=0.85\linewidth]{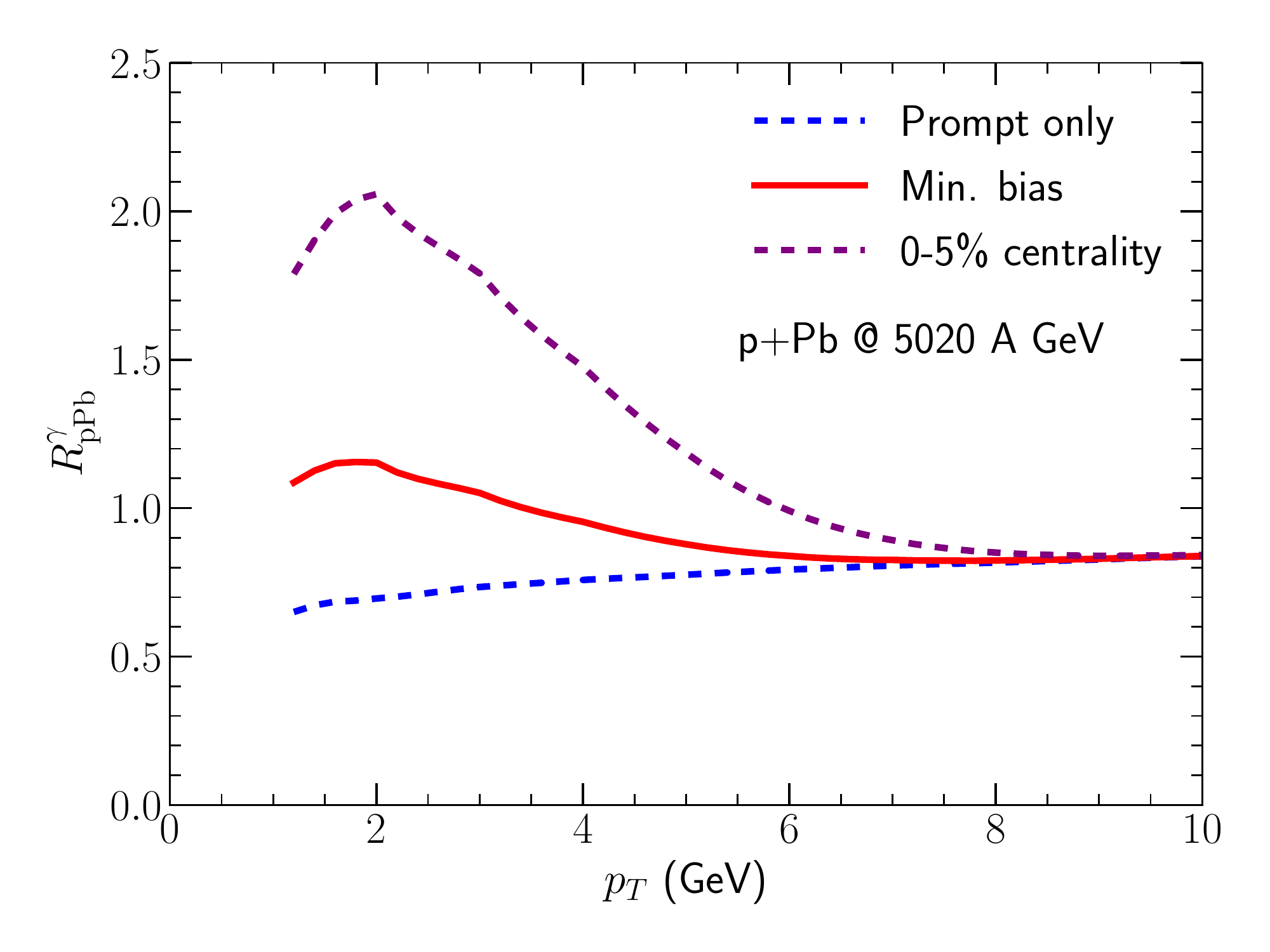}
	\caption{(Color online) $R^\gamma_{pPb}$ of direct photons in p+Pb collisions at 5020 A GeV. The lower line includes only prompt photons, with isospin effect and nCTEQ15 nuclear pdf's as described in Section~\ref{results:photons}. The middle and top lines include medium photons, for minimum bias collisions and 0-5\% central collisions, respectively.}
	\label{fig:RAA_pPb}
\end{figure}

 Finally, Figures~\ref{fig:RAA_OO} and \ref{fig:RAA_pPb} highlight  the direct photon nuclear modification factor as a function of $p_T$ for O+O and p+Pb collisions, respectively, showing separately the effect of the prompt photons. The direct photon nuclear modification factors are shown for 0-5\% central collisions and for the case where one accepts all centralities (minimum bias, 0-100\%). 
The central collisions show a significantly larger direct photon enhancement compared to the minimum bias case for both O+O and p+Pb collisions. In O+O collisions there is a 50\% enhancement in minimum bias collisions (compared to the p+p case), while in p+Pb collisions the enhancement is significantly smaller.  

We find a quantitatively similar thermal photon enhancement as those reported in Refs.~\cite{Shen:2015qba, Shen:2016zpp}, despite the use of a different type of initial state model (Refs. \cite{Shen:2015qba, Shen:2016zpp} used a Glauber model initial state) . The enhancement of direct photon $R^\gamma_{pPb}$ for $p_T < 5$ GeV seen here is an experimental signature of a nearly-thermalized QGP produced in p+Pb collisions. Without thermal radiation contributions, the direct photon $R^\gamma_{AA}$ is expected to be smaller than 1 because of the nuclear shadowing effects. Note that a photon nuclear modification factor,  $R_{pA}^\gamma$ was measured in p+Au collisions at RHIC. That data has large uncertainties, but is nevertheless very suggestive of an enhancement like that described here \cite{pAPhotons}. 

\section{Conclusion}\label{sec:conc}

In this work, we studied simultaneously hadron and photon production in a state-of-the-art multistage model of heavy ion collisions, in a range of different systems at RHIC and LHC.
We studied the effect of varying the pre-equilibrium evolution and chemical equilibration time.

The multistage model included IP-Glasma and \kompost to initialize the hydrodynamic simulation \textsc{music}, which was then followed by the hadronic transport UrQMD.
The introduction of \kompost changes the entropy production of the multistage model as well as the transport coefficients extracted from measurements.
Once these are accounted for,
the influence of the pre-equilibrium model \kompost on the hadronic observables was quantitatively modest,  with its largest effect being on the average transverse momentum of charged hadrons at high centrality (peripheral events). 
In this work both $\eta/s$ and $\zeta/s$ were chosen to best model the hadronic final state data, and only $\zeta/s$ was modified when including the \kompost phase, while we chose $\eta/s$ to be fixed; to explore more detailed effects on all transport coefficients, pre-hydrodynamics can be included in future global Bayesian analyses of relativistic nuclear collisions \cite{Pratt:2015zsa,Bernhard:2019bmu,Nijs:2020roc,Everett:2020xug}.

We further studied the sensitivity of photons to initial deviations from chemical equilibrium.
We introduced a suppression parameter  estimated from QCD kinetic theory studies.
The effect of chemical equilibrium was found to be strongest for the spectra around $p_T\approx 4\,{\rm GeV}$ and for the $v_n$ for $p_T\gtrsim 1-1.5\,{\rm GeV}$ for heavy ion collisions at both RHIC and LHC energies.
The current photon data sets cannot resolve the specific value of the chemical equilibration time, even though the general trend observed for photon elliptic flow seems to favor a larger one.

Much has been written about ``the photon $v_2$ puzzle'' \cite{[{See, for example, }][{, and references therein.}]Gale:2012xq,David:2019wpt}. In a nutshell, the puzzle stems from the fact that the measured direct photon elliptic flow has been found to be as large as that of hadrons, in the region $p_T < 4$ GeV/c, as is also clear from the data shown in this paper. The majority of theoretical models currently underpredict the photon spectrum and elliptic flow. No approach with realistic dynamics can both reproduce photon spectrum and elliptic flow, and this situation has not been modified because of the inclusion of a pre-hydrodynamic phase like \kompostend.

For small collision systems, such as O+O and p+Pb, we predict a strong enhancement of photon production for transverse momenta $p_T\lesssim 4\,{\rm GeV}$, compared to the expectation from prompt photon production. Consequently, experimental measurements of photon spectra in small systems will be essential to determine how small a QGP can be created in nuclear collisions.
 
We emphasize again that photons (real and virtual) enjoy a special status in efforts to characterize strongly interacting matter under extreme conditions. They can be both {\it penetrating} and {\it soft} (on typical strong interaction energy scales): hadronic observables that are penetrating are not soft, and conversely.  Measured together, strong and electroweak particles are a prime example of multi-messenger heavy-ion physics.


\acknowledgments{This work was supported in part by the Natural Sciences and Engineering Research Council of Canada,  in part by the US Department of Energy under Contracts DE-SC0012704, DE-SC0013460,  DE-FG02-05ER41367, DE-AC02-05CH11231, and in part by the National Science Foundation (NSF) under grant number PHY-2012922. This research used resources of the National Energy Research Scientific Computing Center, which is supported by the Office of Science of the U.S. Department of Energy under Contract No. DE-AC02-05CH11231. }



\bibliography{references}

\end{document}